\documentclass[a4paper,11pt]{article}
\pdfoutput=1 % if your are submitting a pdflatex (i.e. if you have
\usepackage{amsmath} 
\allowdisplaybreaks
\usepackage{jheppub} % for details on the use of the package, please
\usepackage{empheq}
\usepackage{bm}
\usepackage{slashed}
\usepackage[T1]{fontenc} % if needed
\usepackage[all]{xy}
\usepackage{rotating}
\usepackage{mathtools}
\usepackage{bbm}% mathbb
\usepackage{dcolumn}
\usepackage{multirow}
\usepackage{mathrsfs}% mathscr
\usepackage{arydshln}% dashed line
\usepackage{verbatim}% comment
\usepackage{graphics}
\usepackage{longtable}

\usepackage{slashed}

\def\a{\alpha}
\def\b{\beta}

\def\CA{{\cal A}}

\def\CD{{\cal D}}

\def\CF{{\cal F}}

\def\CH{{\cal H}}
\def\CI{{\cal I}}

\def\CK{{\cal K}}

\def\CM{{\cal M}}
\def\CN{{\cal N}}
\def\CO{{\cal O}}
\def\CP{{\cal P}}

\def\CR{{\cal R}}
\def\CS{{\cal S}}
\def\CT{{\cal T}}

\def\CW{{\cal W}}

\def\CZ{{\cal Z}}

\def\beq#1\eeq{\begin{align}#1\end{align}}

\newcommand{\be}{\begin{equation}}
	\newcommand{\ee}{\end{equation}}
\newcommand{\ba}{\begin{align}}
	\newcommand{\ea}{\end{align}}%Verbatim cannot recognize this command. 

\newcommand{\bi}{\begin{itemize}}
	\newcommand{\ei}{\end{itemize}}
\newcommand{\braket}[2]{ \langle {#1} | {#2} \rangle}
\makeatletter
\newcommand*{\rom}[1]{\expandafter\romannumeral #1}
\makeatother%For Roman Numeral

\title{\boldmath Non-hyperbolic 3-manifolds and 3D field theories for 2D Virasoro minimal models}

\abstract{Using 3D-3D correspondence, we construct 3D dual bulk field theories for general Virasoro minimal models $M(P,Q)$. These theories correspond to Seifert fiber spaces $S^2 ((P,P-R),(Q,S),(3,1))$ with two integers $(R,S)$ satisfying $PS-QR =1$. In the unitary case, where $|P-Q|=1$, the bulk theory has a mass gap and flows to a unitary topological field theory (TQFT) in the IR, which is expected to support the chiral Virasoro minimal model at the boundary under an appropriate boundary condition. For the non-unitary case, where $|P-Q|>1$, the bulk theory flows to a 3D $\mathcal{N}=4$ rank-0 superconformal field theory, whose topologically twisted theory supports the chiral minimal model at the boundary. We also provide a concrete field theory description of the 3D bulk theory using $T[SU(2)]$ theories. Our proposals are supported by various consistency checks using 3D-3D relations and direct computations of various partition functions.
}

\author{Dongmin Gang,} 
\author{Heesu Kang,}
\author{Seongmin Kim}

\affiliation{
	Department of Physics and Astronomy $\&$ Center for Theoretical Physics,
	\\
	Seoul National University, 1 Gwanak-ro, Seoul 08826, Korea}

\emailAdd{arima275@snu.ac.kr}
\emailAdd{heesu0434@snu.ac.kr}
\emailAdd{seongmin0708@snu.ac.kr}

\begin{document} 
	\maketitle
	\flushbottom

\section{Introduction}
Chiral algebras, also known as vertex algebras, are a powerful and versatile tool in modern theoretical physics. They provide a unifying algebraic framework for understanding diverse physical phenomena, ranging from two-dimensional conformal field theories to the intricate structures in higher-dimensional ($D\geq 3$) topological or supersymmetric quantum field theories \cite{Beem:2013sza,Beem:2014kka,Feigin:2018bkf,Cheng:2018vpl,Costello:2018fnz,Costello:2020ndc,Cheng:2022rqr}.  
Among chiral algebras, rational chiral algebras have rich and rigid mathematical structures and broad applications in physics. They allow only a finite number of irreducible representations, whose characters form vector-valued modular forms. They are closely related to 3D topological field theories via the so-called bulk-boundary correspondence, which provides knot or 3-manifold invariants and describes the universal behaviors of topological phases. The rigid mathematical structures greatly simplify the classification program of rational chiral algebras, and several important classes have been classified. The most famous and successful class is the Virasoro minimal models $M(P,Q)$ \cite{Belavin:1984vu}, which describe universal features of critical phenomena in various 2D systems, such as the Ising model.

In this paper, we study the 3D bulk theories related to the Virasoro minimal models via the bulk-boundary correspondence. In the case of unitary rational chiral algebras, the bulk theories have a mass gap, and the infrared (IR) physics is described by unitary topological field theories. The IR topological quantum field theories (TQFTs) share common modular tensor category structures with the corresponding boundary rational chiral algebras \cite{Witten:1988hf,Moore:1989vd,turaev:1992modular}. For non-unitary rational chiral algebras, recent studies have shown that the bulk theories can be described by topologically twisted theories of an exotic class of superconformal field theories (SCFTs) called 3D $\CN=4$ rank-0 SCFTs \cite{Gang:2018huc,Dedushenko:2018bpp,Gang:2021hrd,Gang:2022kpe,Gang:2023rei,Ferrari:2023fez,Gang:2023ggt,Dedushenko:2023cvd,Baek:2024tuo}. Here, 'rank-0' denotes the absence of Coulomb and Higgs branch operators in the theory. This exotic property proves crucial in realizing rational chiral algebra at the boundary \cite{Costello:2018swh,Beem:2023dub,Ferrari:2023fez}. Our approach begins by realizing the 3D bulk theories for minimal models $M(P,Q)$ through the 3D-3D correspondence \cite{Terashima:2011qi,Dimofte:2011ju,Gang:2018wek} with Seifert fiber spaces, as depicted below:
\begin{align}
\begin{split}
	&\fbox{Seifert fiber spaces} \xrightarrow{  \rm 3D-3D\;   } \fbox{3D theories (gapped theories or $\CN=4$ rank-0 SCFTs)} 
	\\
	& \xrightarrow{  \textrm{bulk-boundary}} \fbox{2 (unitary or non-unitary) minimal models} \label{non-unitary bulk-boundary}
\end{split}
\end{align}
For details of the proposal, refer to \eqref{def: T(P,Q)} and \eqref{T(PQ) from SFS}. Furthermore, we provide a field-theoretic depiction \eqref{field theory for T(P,Q)} of these 3D theories by utilizing the topological structures of the Seifert fiber spaces. Our construction provides a unified framework for the bulk duals of both unitary and non-unitary minimal models.

The rest  of this paper is organized as follows. In Section \ref{sec : MM from 3-manifolds}, we introduce the Virasoro minimal model $M(P,Q)$ and its 3D dual theory 
$\mathcal{T}_{(P,Q)}$, along with the basic dictionaries of the bulk-boundary correspondence. We then propose that the theories $\mathcal{T}_{(P,Q)}$ can be realized as 3D class R theories associated with Seifert fiber spaces (see \eqref{T(PQ) from SFS}). This proposal is tested against various non-trivial 3D-3D relations and bulk-boundary dictionaries, as summarized in Table \ref{table : dictionaries}. In Section \ref{sec : field theory for T(P,Q)}, we provide an explicit field theory description for $\mathcal{T}_{(P,Q)}$ (see \eqref{field theory for T(P,Q)}) along with several non-trivial consistency checks. The Appendices contain technical details of the supersymmetric partition function computations and the identification of decoupled topological field theories from 1-form symmetry 't Hooft anomalies.

\section{Virasoro minimal models from 3-manifolds} \label{sec : MM from 3-manifolds}
In this section, we begin by reviewing basic aspects of Virasoro minimal models. Then, we introduce the bulk-boundary correspondence and the 3D-3D correspondence, summarized in Table \ref{table : dictionaries}. Utilizing these correspondences as guidelines, we propose the 3D bulk duals of minimal models as 3D class R theories associated with 3-manifolds known as Seifert fiber spaces, as given in \eqref{T(PQ) from SFS}.
\subsection{Virasoro minimal model $M(P,Q)$} The minimal model $M(P,Q)= M(Q,P)$ is labeled by two integers, $P$ and $Q$, subject to the following conditions:
\begin{align}
 P, Q  \geq 2 \textrm{ and }  \textrm{gcd}(P,Q)=1\;.
\end{align}
The underlying chiral algebra is the Virasoro algebra, with the 2D central charge given by:
\begin{align}
c_{2d}= 1- \frac{6(P-Q)^2}{P Q}\;. \label{central charge}
\end{align}
The model includes critical Ising CFT $M(3,4)$, tricritical Ising CFT $M(4,5)$ and Lee-Yang CFT $M(2,5)$.  The 2D RCFT can be unitary or non-unitary, depending on $(P,Q)$:
\begin{align}
M(P,Q) \textrm{ is } 
\begin{cases}
\textrm{unitary},  &\; \textrm{if } |P-Q|=1
\\
\textrm{non-unitary}, & \; \textrm{otherwise}\;.
\end{cases}
\end{align}
There are $N_{(P,Q)}:=\frac{(P-1)(Q-1)}2$ primaries $\CO_{(a,b)}$  labeled by two integers $1 \leq a <P$ and $1 \leq b <Q$ modulo an equivalence relation $\CO_{(a,b)} = \CO_{(P-a,Q-b)}$. The conformal dimensions $h$ of the primaries are given by:
\begin{equation}
h_{(a,b)}=\frac{(Pb-Qa)^2-(P-Q)^2}{4PQ}\;, \label{conformal dimensions 1}
\end{equation}
and the conformal characters are:
\begin{align}
\chi_{(a,b)} (q) = \frac{q^{h_{(a,b)} - \frac{c_{2d}}{24}}}{(q)_\infty } \sum_{n\in \mathbb{Z}} \left(q^{n^2 P Q+ n (Qa-Pb)} - q^{(nP+a)(nQ+b)} \right)\;.
\end{align}
Here $(q)_\infty = \prod_{k=1}^\infty (1-q^k)$ as usual. Under the S-transformation, $q :=e^{2\pi i \tau}\rightarrow \tilde{q}:=e^{2\pi i (- \frac{1}\tau)}$,  the characters transform as:
\begin{align}
\chi_{\alpha} (\tilde{q}) = \sum_{\beta} S_{\a \b} \chi_\b (q)\;, \label{modular S-matrix 1}
\end{align}
with the modular $S$-matrix given by:
\begin{equation}
S_{(a,b),(a',b')}=-\sqrt{\frac{8}{PQ}} \ (-1)^{ba'+b'a \ }\text{sin}(\pi \frac{P}{Q}bb') \ \text{sin}(\pi \frac{Q}{P}a a')\;. \label{modular S-matrix 2}
\end{equation}
In the above, $\alpha, \beta =0,1, \ldots, N_{(P,Q)} -1$ are  collective indices for the primaries, where  $\alpha=0$ corresponds to the vacuum module, i.e., $\alpha =0 \leftrightarrow (a,b)=(1,1)$. 
\subsection{Bulk dual 3D $\CT_{(P,Q)}$ theory} 
The bulk-boundary correspondence relates a 3D bulk (semi-simple and finite) topological field theory $\CT$ to 2D chiral rational conformal field theories (RCFTs).  The boundary chiral RCFT $\chi R$ depends on the choice of  holomorphic boundary condition $\mathbb{B}$, and  the corresponding RCFT will be denoted as $\chi R[\CT,\mathbb{B}]$:
\begin{align}
(\textrm{3D topological field theory } \CT) \xrightarrow{\;\textrm{at boundary with $\mathbb{B}$} \;} \textrm{(2D chiral RCFT $\chi R[\mathcal{T}, \mathbb{B}]$)}\;.
\end{align}
To realize the full (diagonal) RCFT $\CR$, one needs to put the bulk theory on an interval, $\CM_2 \times [0,1]$, with 
 the holomorphic boundary condition $\mathbb{B}$ on both boundaries. In the IR, the system  flows to the 2D RCFT $\CR[\CT, \mathbb{B}]$ on $\CM_2$. 

 For unitary case, the bulk theory is a unitary TQFT, which describes the universal IR behavior of a (2+1)D gapped system,  such as fractional quantum Hall system.  
 For non-unitary case, recent studies show that the bulk theories can be described by topologically twisted theories of an exotic class of superconformal field theories called 3D 
 $\CN=4$ rank-0 SCFTs.  The bulk-boundary correspondence for the non-unitary case can be summarized as follows:
 \begin{align}
 \begin{split}
 & \textrm{a 3D $\CN=4$ rank-0 SCFT } \CT \xrightarrow{\;\textrm{a top'l twisting}\;} \textrm{non-unitary TQFT } \CT^{\rm top}
 \\
 & \xrightarrow{\textrm{ at boundary with $\mathbb{B}$ }}  \textrm{non-unitary chiral RCFT $\chi \CR[\CT, \mathbb{B}]$}\;.
 \end{split}
 \end{align}
 Rank-0 means there are no Coulomb and Higgs branch operators in the theory.  The exotic property turns out to be crucial to realize rational chiral algebra at the boundary after a topological twisting. 
 There are two possible choices of topological twistings (`top'=$A$ or $B$) denoted as $A$ and $B$ twistings. 
 
  Let $\CT_{(P,Q)}$  be the bulk dual theory  related to the chiral minimal model $\chi M(P,Q)$ via the bulk-boundary correspondence:
 \begin{align}
 \begin{split}
 &\textbf{Def : 3D $\CT_{(P,Q)}$ theory is defined as  }
 \\
 & \textrm{For }|P-Q|=1,\; \CT_{(P,Q)}  \textrm{ is a 3D unitary TQFT with}
 \\
 &\CT_{(P,Q)}\xrightarrow{\textrm{ at boundary with  a proper $\mathbb{B}$ } } \chi M(P,Q)\;,
 \\
 &\textrm{For }|P-Q|>1, \; \CT_{(P,Q)}  \textrm{ is a $\CN=4$ 3D rank-0 SCFT with}
 \\
 & \CT_{(P,Q)} \xrightarrow{\textrm{a top'l twisting}\;} \textrm{non-unitary TQFT } \CT^{\rm top}_{(P,Q)} \xrightarrow{\textrm{ at boundary with a proper $\mathbb{B}$ } } \chi M(P,Q)\;.
 \end{split} \label{def: T(P,Q)}
 \end{align}
 The main goal of this paper is to construct the bulk theory $\CT_{(P,Q)}$ dual to the minimal model $M(P,Q)$.
 
 Basic dictionaries of the bulk-boundary correspondence are summarized in the first and second column of Table \ref{table : dictionaries}. Refer to \cite{Gang:2021hrd,Gang:2023rei} for details. In the table $T_{\rm irred}[M]$ corresponds to the bulk theory $\CT$ for a closed 3-manifold $M$. We will realize the bulk theory as an IR fixed point of 3D $\CN\geq 2$ supersymmetric gauge theories, and the bulk quantities in the table are related to the partition functions on various symmetric backgrounds, which are RG-invariant. When the SUSY background is $\CM_{g,p}$, a degree $p$ circle bundle over a genus $g$ Riemann surface, the partition function can be given in the following form \cite{Benini:2015noa,Benini:2016hjo,Closset:2016arn,Closset:2018ghr}
 \begin{align}
 \CZ^{\CM_{g,p}} = \sum_{\vec{x}_\alpha \in \CS_{\rm BE}} \CH(\vec{x}_\alpha)^{g-1} \CF(\vec{x}_\alpha)^p\;. \label{twisted ptns}
 \end{align} 
 Here $\vec{x}_\alpha \in \CS_{\rm BE}$ are called Bethe-vacua, which are ground states on a two-torus $\mathbb{T}^2$ when the bulk theory is a topological field theory. $\CH$ and $\CF$ are called `handle-gluing' and `fibering' operators, respectively. Let $\CZ^{S^3_b}$ denotes the supersymmetric partition function on a squashed 3-sphere $S^3_b$ \cite{Kapustin:2009kz,Jafferis:2010un,Hama:2010av}:
 \begin{align}
 S^3_b := \{(z,w)\in \mathbb{C}^2\;:\; b^2 |z|^2+ b^{-2} |w|^2=1 \}\;. \label{squashed 3-sphere}
 \end{align}
When the bulk theory flows to a 3D $\CN=4$ rank-0 SCFT, the SUSY partition functions have non-trivial dependence on two real parameters $(m,\nu)$. Here $m$ denotes a real mass for the $U(1)_A$ flavor symmetry, whose charge $A$ is given in terms of an $\CN=2$ subalgebra by
\begin{align}
A = J^C_3 - J^H_3\;.  \label{charge A}
\end{align}
$J^C_3$ and $J^H_3$ are two Cartans of $SU(2)^{C} \times SU(2)^H \simeq SO(4)$ R-symmetry of the $\CN=4$ theory, which are normalized as $J^{C/H}_3 \in \mathbb{Z}/2$. $\nu$ parametrizes the mixing between the $U(1)_R$ symmetry of the $\CN=2$ subalgebra and the $U(1)_A$ symmetry as follows:
\begin{align}
R_\nu = R_{\nu=0} + \nu A=\left(J^C_3+J^3_H\right) + \nu \left(J^C_3-J^3_H\right) \;. \label{R-symmetry mixing}
\end{align}
%
\begin{comment} 
 %
When the bulk theory flows to a 3D $\CN=4$ rank-0 SCFT, the SUSY partition functions have non-trivial dependence on $(M,\nu)$. In terms of an $\CN=2$ subalgebra, the $\CN=4$ theory has a flavor symmetry $U(1)_A$ whose charge $A$ is given by:
%
\begin{align}
A = J^C_3 - J^H_3\;,  \label{charge A}
\end{align}
%
where  $J^C_3$ and $J^H_3$ are two Cartans of $SU(2)^{C} \times SU(2)^H \simeq SO(4)$ R-symmetry of the $\CN=4$ theory. They are normalized as $J^{C/H}_3 \in \mathbb{Z}/2$. $M$ denotes the properly rescaled real mass of the $U(1)_A$ symmetry. $\nu$ parametrizes the mixing between the $U(1)_R$ symmetry of the $\CN=2$ subalgebra and the $U(1)_A$ symmetry as follows:
%
\begin{align}
R_\nu = R_{\nu=0} + \nu A=\left(J^C_3+J^3_H\right) + \nu \left(J^C_3-J^3_H\right) \;. \label{R-symmetry mixing}
\end{align}
%
\end{comment}
For rank-0 SCFTs, the supersymmetric partition functions in the following limits:
\begin{align}
\begin{split}
&\textrm{A-twisting limit : } (m, \nu) \rightarrow (0, \nu=-1)\;,
\\
&\textrm{B-twisting limit : } (m, \nu) \rightarrow (0, \nu=1)\;, \label{A/B twisting limit}
\end{split}
\end{align}
are known to compute the partition functions of topologically A-twisted or B-twisted theory, respectively. In the limits, the squashed 3-sphere function becomes independent on the squashing parameter modulo a phase factor in \eqref{phase factor ambiguity of S3b}. 
Throughout the paper, we will consider topologically A-twisted theories. In the table, we define 
\begin{align}
\begin{split}
&\{ \CF^{top}, \CH^{top}, \CZ_b^{top} \}:=\{ \CF , \CH , \CZ^{S^3_b} \}|_{(m, \nu)\rightarrow (0, -1)} \;.%,
%\\
%&\{ \CF^{B}, \CH^B, \CZ_b^B \}:=\{ \CF , \CH , \CZ^{S^3_b} \}|_{(m, %\nu)\rightarrow (0, +1)} \;.
\end{split}
\end{align} 
On the hand, the SUSY partition functions of a 3D $\CN=4$ rank-0 SCFT at $(m=0, \nu=0)$ compute the partition functions at the superconformal point, and we define:
\begin{align}
\begin{split}
 \CZ_b^{\rm con}:=\CZ^{S^3_b}|_{(m, \nu)\rightarrow (0, 0)} \;. \label{conformal limit}
\end{split}
\end{align} 
When the 3D gauge theory has a mass gap and flows to a unitary TQFT in the IR, the SUSY partition functions are independent on the $(m,\nu)$ and its squashed 3-sphere partition function is independent on the $b$ modulo a phase factor in \eqref{phase factor ambiguity of S3b}.

\subsection{$\CT_{(P,Q)}$ from non-hyperbolic 3-manifolds}
Let $T_{\rm irred}[M]$ denote the 3D class R theory associated with a closed 3-manifold $M$, whose field theory description is proposed in  \cite{Dimofte:2011ju,Gang:2018wek}.  The theory is believed to describe an effective 3D field theory of 6D $A_1$ $\CN =(2,0)$ superconformal field theory compactified on the 3-manifold $M$.   The subscript `irred'  emphasizes that the theory only sees an irreducible component of $SL(2,\mathbb{C})$  flat connections on $M$ rather than all the flat connections \cite{Chung:2014qpa}.  The theory should be distinguished from $T_{\rm full}[M]$ studied in \cite{Gadde:2013sca,Gukov:2016gkn,Gukov:2017kmk,Cheng:2018vpl,Eckhard:2019jgg,Chung:2019khu,Assel:2022row,Chung:2023qth},  which is assumed to see all flat connections. Unlike $T_{\rm irred}$, however,  there is no known systematic algorithm for constructing the field theory description of $T_{\rm full}[M]$ for a general 3-manifold $M$. It is even uncertain whether the theory $T_{\rm full}[M]$  exists as a genuine 3D field theory for general $M$ \cite{Gang:2018wek}.

\begin{figure}[h]
	\centering
	\includegraphics[width=0.3\textwidth]{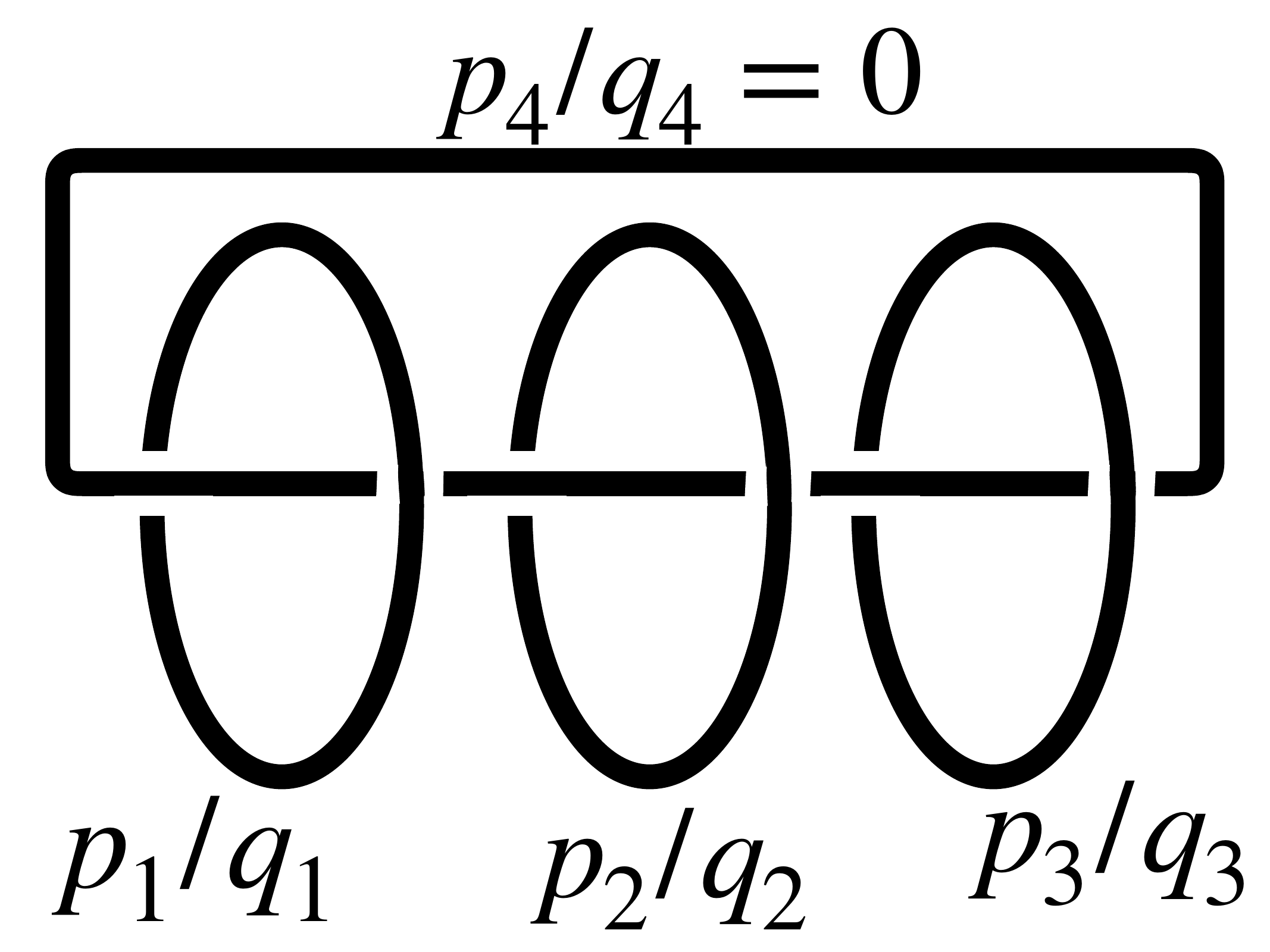}
	\caption{A Dehn surgery representation of   $S^2 (\vec{p},\vec{q}):=S^2 ((p_1, q_1),(p_2,q_2),(p_3, q_3))$. }
	\label{fig:SFS}
\end{figure}

For Seifert fiber spaces (SFSs) $M=S^2 (\vec{p},\vec{q})$ in Figure \ref{fig:SFS},  
the IR phases of the 3D field theory $T_{\rm irred}[M]$ have been analyzed in \cite{Choi:2022dju}. It was empirically found that
\begin{align}
\begin{split}
&T_{\rm irred}[M = S^2 ((p_1, q_1),(p_2,q_2),(p_3,q_3))] 
\\
&\xrightarrow{\textrm{ In the IR }}\begin{cases}
\textrm{a unitary TQFT}, \;\; & q_i = \pm 1 \;(\textrm{mod }p_i) \; \forall i=1,2,3\;,
\\
\textrm{a rank-0 SCFT}, \;\; &\textrm{otherwise}
\end{cases}
\end{split} \label{IR phases of T[SFS]}
\end{align}
By combining the 3D-3D correspondence for the Seifert fiber spaces and the bulk-boundary correspondence, one can consider following correspondence:
\begin{align}
M=S^2 (\vec{p},\vec{q}) \xrightarrow{\textrm{ 3D-3D correspondence }}T_{\rm irred}[M] \xrightarrow{\textrm{ Bulk-boundary }} \textrm{2D chiral RCFT } \chi\mathcal{R}[M;\mathbb{B}]\;.
\end{align}
Basic dictionaries for the correspondence are summarized in Table \ref{table : dictionaries}.
%The Seifert fibered space $S^2 ((p_1, q_1),(p_2,q_2),(p_3, q_3))$ can be represented by a Dehn surgery as depicted in Figure \ref{fig:SFS}.
%
%
\begin{table}[h]
	\begin{center}
		\begin{tabular}{ |c|c|c|} 
			\hline
			${\rm2D\;  chiral\; RCFT \;}\chi\mathcal{R}[M;\mathbb{B}]$& $T_{\rm irred}[M]$ & $SL(2,\mathbb{C})$ CS on $M$
			\\
			\hline
			(unitary)/(non-unitary) & (mass gap)/(rank-0 SCFT) & \textrm{equation} \eqref{IR phases from 3-manifolds}
			\\
			\hline
			Primary $\CO_{\alpha=0,\ldots, N-1}$& Bethe-vacuum $\vec{x}_\alpha \in \CS_{\rm BE}$  \quad &   $\rho_\alpha \in \chi_{\rm irred}[M]$ \\ 
				\hline
			Conformal dimension $h_\alpha $&  $e^{2\pi i h_\alpha} = \CF^{\rm top}(\vec{x}_\alpha)/\CF^{\rm top}(\vec{x}_{\a=0})$  & $e^{2\pi i h_\alpha} =e^{2\pi i \left( {\rm CS}[\rho_{\a=0}] -{\rm CS}[\rho_{\a}]\right) }$  \\ 
			\hline
			$S_{0 \alpha}^2$  & $(\CH^{\rm top}(\vec{x}_\alpha))^{-1}$ & $1/(2 \textrm{Tor}[\rho_\alpha])$  \\ 
				\hline
			$|S_{0 0}|$  & $|\CZ^{\rm top}_b|$  & $\big{|}\sum_{\rho_\alpha }\frac{e^{-2\pi i {\rm CS}[\rho_\alpha]}}{2 \textrm{Tor}[\rho_{\alpha}]} \big{|}$  \\ 
				\hline
			$\textrm{min}_\alpha \{|S_{0 \alpha}|\}$  & $e^{-F}=|\CZ^{\textrm{con}}_{b=1} | $& $\textrm{min}_\alpha  \{ 1/\sqrt{2 \textrm{Tor}[\rho_\alpha]}\}$  \\ 
			\hline 
		\end{tabular}
		\caption{Basic dictionaries for the correspondence among non-hyperbolic 3-manifolds $M$, 3D bulk theories $T_{\rm irred}[M]$, and  2D chiral  RCFTs $\chi\mathcal{R}[M]$ for $M=S^2 ((\vec{p},\vec{q}))$ with $H^1 (M, \mathbb{Z}_2)=1$. The superscripts `top' and `con' denote the partition function in $A$-twisting limit \eqref{A/B twisting limit} and at the superconformal point \eqref{conformal limit} of rank-0 SCFT respectively. For the mass gap case, the superscripts can be ignored. The $S_{\alpha \beta} $ is the modular S-matrix in \eqref{modular S-matrix 1} and \eqref{modular S-matrix 2}. $\CH$ and $\CF$ are handle gluing and fibering operators appearing in the twisted partition function \eqref{twisted ptns}. $
		\CZ_b$ is the squashed 3-sphere partition function.  $\textrm{CS}[\rho]$ and $\textrm{Tor}[\rho]$ denotes the Chern-Simons invariant and adjoint Reidemeister torsion of an irreducible flat connection $\rho$.}   \label{table : dictionaries}
	\end{center}
\end{table}
Let us briefly explain the basic 3D-3D dictionaries in the table. Refer to \cite{Gang:2018hjd,Gang:2019uay,Benini:2019dyp,Cho:2020ljj,Cui:2021lyi,Bonetti:2024cvq} for details.  The dictionary is valid only for the 3-manifold with trivial $H^1(M, \mathbb{Z}_2)$. 
The $\chi_{\rm irred}(M)$ in the table is  the set of  (adjoint-)irreducible $SL(2,\mathbb{C})$ characters on a 3-manifold $M$, which is defined as:
\begin{align}
\begin{split}
\chi_{\rm irred}[M] &= \{\rho\in \textrm{Hom}\left[\pi_1(M)\rightarrow SL(2,\mathbb{C})] \;:\; \textrm{$\textrm{dim}H(\rho) = 0$}\right\}/\sim\;,
\\
&\qquad  \textrm{where } H(\rho):=\{g \in SL(2,\mathbb{C})\;:\;[g,\rho(a)]=0 \;\;\forall a \in \pi_1 (M)\}\;.
\end{split}
\end{align}
The equivalence relation is defined as\footnote{In \cite{Cho:2020ljj}, they consider equivalence up to  $SL(2,\mathbb{C})$ conjugation. But the trace equivalence relation $\sim$, which is a weaker equivalence, turns out to be more relevant in the 3D-3D correspondence \cite{Cui:2021lyi}.    }
\begin{align}
\rho_1 \sim \rho_2 \textrm{ if }\textrm{Tr}(\rho_1 (a)) = \textrm{Tr}(\rho_2 (a))\;,  \;\;\forall a \in \pi_1 (M)\;.
\end{align}
The condition $\textrm{dim} H(\rho) =0$ corresponds to the irreducibility of the homomorphism $\rho$. A  homomorphism $\rho$ defines an $SL(2,\mathbb{C})$ flat connection $\CA_\rho$, $d\CA_\rho +\CA_\rho \wedge \CA_\rho=0$.
The ${\rm CS}[\rho]$ and $\textrm{Tor}[\rho]$ are basic topological invariants of the flat connection called Chern-Simons invariant and the adjoint Reidemeister torsion, respectively. The CS invariant is defined as 
\begin{align}
{\rm CS}[\rho]:=\frac{1}{8\pi^2}  \int \textrm{Tr}\left(\CA_\rho d \CA_\rho +\frac{2}3 \CA_{\rho}^3\right)\;\; \textrm{ (mod $1$) }.
\end{align}
The adjoint torsion $\textrm{Tor}[\rho]$ appears as the 1-loop part of perturbative expansion of $SL(2,\mathbb{C})$ CS theory around the flat connection $\CA_\rho$ \cite{Witten:1988hf}. For $M=S^2 (\vec{p},\vec{q})$ with trivial $H^1(M, \mathbb{Z}_2)$, one can determine the IR phase of the $T_{\rm irred}[M]$ theory in the following way \cite{Cho:2020ljj,Choi:2022dju}
\begin{align}
\begin{split}
&T_{\rm irred}[M=S^2 (\vec{p},\vec{q})] 
\\
&\xrightarrow{\; \textrm{in the IR}\;}  \begin{cases}  \textrm{unitary TQFT}\;, &  \textrm{if } |\sum_{\rho \in \chi} \frac{e^{-2\pi i {\rm CS}[\rho_\alpha]}}{2 \textrm{Tor}[\rho]}| \leq  \frac{1}{\sqrt{|2 \textrm{Tor}[\rho_\alpha]|}} \;\;\forall \;\rho_\a \in \chi
\\
  \textrm{3D  rank-0 SCFT} \;, &\textrm{otherwise }
\end{cases} 
\end{split}\label{IR phases from 3-manifolds}
\end{align}
Here, $\chi$ abbreviates $\chi_{\text{irred}}[M]$.

\textbf{As  the main result of this section}, we propose that
\begin{align}
\begin{split}
&\CT_{(P,Q)} \simeq  T_{\rm irred}\left[S^2((P,P-R),(Q,S), (3,1))\right]\;. \label{T(PQ) from SFS}
\end{split}
\end{align}
The 3D theory $\CT_{(P,Q)}$ is defined as the bulk dual of the Virasoro minimal model $M(P,Q)$ as in \eqref{def: T(P,Q)}. Here $(R,S)$ is chosen to satisfy
\begin{align}
\begin{pmatrix} P & Q \\ R & S
\end{pmatrix} \in SL(2,\mathbb{Z})\;.
\end{align}
It  fixes the $(R,S)$ modulo a shift $(R,S)\rightarrow (R,S)+\mathbb{Z}(P,Q)$, and we will claim that the $T_{\rm irred}\left[S^2((P,R-P),(Q,-S), (3,1)) \right]$ is independent of the shift:
\begin{align}
\begin{split}
&T_{\rm irred}\left[S^2((P,P-R),(Q,S), (3,1))  \right] \simeq  T_{\rm irred}\left[S^2((P,P-\tilde{R}),(Q,\tilde{S}), (3,1))  \right] \;
\\
&\textrm{where } (\tilde{R},\tilde{S}) = (R,S) + n(P,Q)\;\textrm{and for arbitrary $n\in \mathbb{Z}$}\;. \label{independence of R and S}
\end{split}
\end{align}
Throughout this paper, we use the following two equivalence relations denoted by $\sim$ and $\simeq$ among 3D gauge theories,
\begin{align}
\begin{split}
&\CT_1 \sim \CT_2 \textrm{ if two theories are IR equivalent up to some `topological operations'}\;,
\\
&\CT_1 \simeq \CT_2 \textrm{ if two theories are IR equivalent up to some `minimal topological operations'}\;. \label{two equivalence relations}
\end{split}
\end{align}
 Topological operations include tensoring with a unitary TQFT, gauging of finite (generalized) symmetries, time-reversal, and so on. On the other hand, the minimal topological operations are topological operations which preserve the absolute values of partition functions on arbitrary closed 3-manifolds.  The minimal ones include tensoring with an invertible TQFT, time-reversal, and so on. Notice that $\simeq $ is a stronger equivalence than  $\sim$.
The $\CT_{(P,Q)}$, like $M(P,Q)$, is  invariant under the exchange of $P \leftrightarrow Q$:
\begin{align}
\begin{split}
\CT_{(Q,P)} &\simeq  T_{\rm irred}\left[S^2((Q,Q-\tilde{R}),(P,\tilde{S}), (3,1))\right] \textrm{ with } \begin{pmatrix} Q & P \\ \tilde{R} & \tilde{S}
\end{pmatrix} \in SL(2,\mathbb{Z})\;.
\\
&\simeq  T_{\rm irred}\left[S^2((P,(\tilde{S}-P)+P), (Q,(Q-\tilde{R})),(3,1))\right]  \simeq  \CT_{(P,Q)}\;.
\end{split}
\end{align}
In the 2nd line, we use the fact  that $\begin{pmatrix} P & Q \\ -\tilde{S}+P & Q-\tilde{R}
\end{pmatrix} \in SL(2,\mathbb{Z})$. 

Let us check the proposal using the dictionaries in Table \ref{table : dictionaries}.  First, note that the 3-manifold $S^2 ((P,P-R),(Q,S),(3,1))$ has trivial $H^1(M,\mathbb{Z}_2)$\footnote{$S^2((\vec{p},\vec{q}))$ has trivial $H^1(M, \mathbb{Z}_2)$ if and only if $p_1 p_2p_3 \left( \frac{q_1}{p_1}+\frac{q_2}{p_2} +\frac{q_3}{p_3}\right) \in 2\mathbb{Z}+1$.} and thus one can use the dictionaries. The fundamental group of the SFS can be presented as:
\begin{equation}
\pi_1(S^2 (\vec{p},\vec{q}))= \braket{x_1,x_2,x_3,h}{x_i^{p_i}h^{q_i}=1,x_1x_2x_3=1, \textrm{$h$ is central}}\;. \label{fund group}
\end{equation}
As studied in \cite{Cui:2021lyi},   irreducible characters on  $M=S^2 (\vec{p},\vec{q})$ can be specified by the quadruple $(n_1,n_2, n_3, \lambda)$ with $n_{k=1,2,3} \in \mathbb{Z}_{\geq 0}/2$ and $\lambda \in \{0, \frac{1}2\}$, where
\begin{align}
\textrm{Tr} (\rho (x_k)) = 2\cos (\frac{2\pi n_k}{p_k})\;, \; \rho (h)= \exp (2\pi i \lambda) \mathbb{I}_2\;.
\end{align}
Note that the $\rho(h)$ should  be an element of center subgroup $\mathbb{Z}_2$ ($\mathbb{I}_2$ or $-\mathbb{I}_2$)  of $SL(2,\mathbb{C})$ in order for $\rho$ to be (adjoint)-irreducible, i.e. $\textrm{dim}H(\rho)=0$. Otherwise, the fundamental group relation can only be met  when $\rho(x_k) \in \{\pm \mathbb{I}_2\}$ for all $k=1,2,3$, and thus $\textrm{dim}H(\rho)>0$. 

Before going into the details of the character variety, let us first check the proposed duality in \eqref{independence of R and S} of R and S at the level of the character variety, which corresponds to the set of  Bethe-vacua in the $T_{\rm irred}[M]$ theory.
One can easily construct a natural one-to-one map between the character varieties with different choices of $(R,S)$ as follows
\begin{align}
\begin{split}
&\rho \in \chi_{\rm irred}\left[S^2 ((P,P-R), (Q,S),(3,1))\right] 
\\
& \xleftrightarrow{\quad \textrm{one-to-one}\quad } \tilde{\rho} \in \chi_{\rm irred}\left[S^2 ((P,P-\tilde{R}), (Q,\tilde{S}),(3,1))\right]_{(\tilde{R},\tilde{S}) = (R,S)+n (P,Q)} \;,
\\
&\textrm{where}
\\
&\tilde{\rho} (x_1) = \rho(x_1) \rho(h)^{n}, \; \;\tilde{\rho} (x_2) = \rho(x_2) \rho (h)^{-n}, \;\; \tilde{\rho}(x_3) = \rho (x_3), \;\; \tilde{\rho}(h) = \rho (h)\;.
\end{split}
\end{align}
Basic invariants, $\textrm{CS}[\rho]$ and $\textrm{Tor}[\rho]$, of  irreducible characters are preserved under the map. 

The character variety is studied in \cite{Cui:2021lyi} and they found that 
\begin{align}
\begin{split}
\chi_{\rm irred}(M) = &\left\{ \left( n_{p_1,q_1}(j_1), n_{p_2,q_2}(j_2), n_{p_3,q_3}(j_3),\frac{1}{2}   \right) \middle| \ j_k \in [0, \ldots, p_k - 2]^{e} \right\} \\
&\bigsqcup \bigg{\{} \left( n_{p_1,q_1}(j_1), n_{p_2,q_2}(j_2), n_{p_3,q_3}(j_3),0  \right) \bigg{|} \ j_k \in [0, \ldots, p_k - 2]^{o} \bigg{\}},
\end{split}
\end{align}
where 
\begin{align}
n_{p,q}(j)= \begin{cases}
\frac{p-j-1}{2}\;, & q \textrm{ and } j \in 2\mathbb{Z}
\\
\frac{j+1}{2}\;, &  \text{otherwise}
\end{cases} \label{npq}
\end{align}
Here $[0,\ldots, p]^{e/o}$ denotes the set of even/odd numbers between $0$ and $p$. So the irreducible  characters on $S^2 (\vec{p},\vec{q})$ are labeled  by $\vec{j} \in \prod_{k=1}^3[0...p_k-2]^e \sqcup \prod_{k=1}^3[0,...,p_k-2]^o$.
In our case, $p_3=3$ and $j_3 \in \{ 0,1 \}$, so  $ (j_1, j_2)\in \prod_{k=1}^2[0,...,p_k-2]^e \sqcup \prod_{k=1}^2[0,...,p_k-2]^o$ wholly fix the $\vec{j}$ as well as $\lambda$. %
In the labeling, the adjoint Reidemeister torsion and Chern-Simons invariant (mod 1) of a $SL(2,\mathbb{C})$ character $\rho_{\vec{j}}$ on  the 3-manifold $S^2((P,P-R),(Q,S), (3,1)) $ are \cite{Cui:2021lyi}
\begin{align}
\begin{split}
&\textrm{Tor}(\rho_{\vec{j}} )=\prod_{k=1}^3 \frac{p_i}{4 \sin^2 (\frac{2\pi r_k n_k}{p_k})} = \frac{PQ}{16\text{sin}^2(\frac{2\pi Q n_1}{P})\text{sin}^2(\frac{2\pi P n_2}{Q})}\;,
\\
&\textrm{CS}(\rho_{\vec{j}}) = \sum_{k=1}^{3}\frac{-c_k}{4p_k}(j_k+1)^2\;.
\end{split}
\end{align}
where $c_k$ is :
\begin{align}
c_k=    \begin{cases}
p_kq_ks_k-r_k\;, & \ q_k\ \text{odd} \\
p_kq_ks_k-r_k(p_k-1)^2, & \ q_k\ \text{even}
\end{cases}
\end{align}
Here $(r_k, s_k) \in \mathbb{Z}^2$ is chosen to satisfy $p_ks_k-q_kr_k=1$, and $n_1 = n_{(p_1, q_1)} (j_1) $ and $n_2:= n_{(p_2, q_2)} (j_2) $ in \eqref{npq}.

We propose the following one-to-one map between the primaries $\CO_{(a,b)}$ of $M(P,Q)$ and the irreducible characters $\rho_{\vec{j}}$ on $S^2((P,P-R),(Q,S), (3,1)) $:
\begin{align}
\begin{split}
&\CO_{(a,b)} \leftrightarrow \rho_{\vec{j}}\;,   \textrm{ where }
\\
%&\textrm{Case I : Neither $P$ nor $Q$ is in $4\mathbb{Z}$}
%\\
%& \quad (a,b)= \begin{cases}
%(j_1+1,j_2+1)\;, & j_1,j_2 \ \text{even}
%\\
%(P-j_1-1,j_2+1)\;, & j_1,j_2 \ \text{odd}
%\end{cases}
%\\
%&\textrm{Case II : One of $P$ and $Q$ is in $4\mathbb{Z}$}
%\\
& \quad i) \;\;(a,b)= \begin{cases}
(j_1+1,j_2+1)\;, & j_1,j_2 \ \text{even}
\\
(P-j_1-1,j_2+1)\;, & j_1,j_2 \ \text{odd} 
\end{cases} \;,
\\
&\quad \textrm{ if $(P,R,S) \in (4\mathbb{Z}, 2\mathbb{Z}+1, 2\mathbb{Z})$ or $(Q, R,S) \in (4\mathbb{Z}, 2\mathbb{Z}+1, 2\mathbb{Z}+1)$}
\\
& \quad ii) \;\; (a,b)= \begin{cases}
(P-j_1-1,j_2+1)\;, & j_1,j_2 \ \text{even}
\\
(j_1+1,j_2+1)\;, & j_1,j_2 \ \text{odd}\;
\end{cases}\;,
\\
&\quad \textrm{otherwise}\;.
\end{split} \label{characters-to-primaries}
\end{align}
Under the map, one can  check that $(2{\rm Tor}(\rho_\alpha))^{-1}$ (resp. $\textrm{CS}[\rho_{\a=0}]-\textrm{CS}[\rho_{\alpha}]$) equals to $S_{0\alpha}^2$   (resp. $h_\alpha$ (mod 1)) of $M(P,Q)$. 
\\
\\
Let us give some concrete examples. 
\paragraph{Example: $M(3,4) =(\textrm{Ising CFT})$ from $ S^2 ((3,4),(4,-1),(3,1))$} We choose $(R,S) = (-1,-1)$. There are  3 irreducible $SL(2,\mathbb{C})$ characters $\rho_{\vec{j}}$:
\begin{table}[h]
	\begin{center}
		\begin{tabular}{|c|c|c|c|c|c|} 
			\hline
			$(j_1, j_2)$ & $(\vec{n},\lambda)$ & $\textrm{CS}[\rho]$ & $\textrm{Tor}[\rho]$ & $\CO_{(a,b)}$ & $\textrm{CS}[\rho_{\a=0}] -\textrm{CS}[\rho]$
			\\
			\hline
			$(0, 0)$ & $(1, \frac{1}2, \frac{1}2, \frac{1}2)$ & $\frac{31}{48}$ & $2$ & $\CO_{(1,1)}=\CO_{(2,3)}$ & $0$
			\\
			\hline
			$(0, 2)$ & $(1, \frac{3}2, \frac{1}2, \frac{1}2)$ & $\frac{7}{48}$ & $2$ & $\CO_{(1,3)} =\CO_{(2,1)}$ & $\frac{1}2$ 
			\\
			\hline
			$(1, 1)$ & $(1, 1, 1, 0)$ & $\frac{7}{12}$ & $1$ & $\CO_{(1,2)} = \CO_{(2,2)}$ & $\frac{1}{16}$
			\\
			\hline
		\end{tabular}
		\caption{Ising CFT  from $S^2 ((3,4),(4,-1),(3,1))$. We use the characters-to-primaries map in \eqref{characters-to-primaries}. The result is compatible with that $S_{(1,1),(1,1)} = S_{(1,1),(2,1)} = \frac{1}2, S_{(1,1),(2,2)} = \frac{1}{\sqrt{2}} $ and $h_{(2,1)}= \frac{1}2, h_{(2,2)}= \frac{1}{16}$. } 
	\end{center}
\end{table}
\\
\\
\paragraph{Example: $M(3,4) =(\textrm{Ising CFT})$ from $ S^2 ((3,1),(4,3),(3,1))$} This time, we choose $(R,S) = (2,3)$. Due to the different choice of $(R,S)$, the correspondence between the primaries and the irreducible characters is different from the case when $(R,S) = (-1,-1)$.:
\\
\begin{table}[h]
	\begin{center}
		\begin{tabular}{|c|c|c|c|c|c|} 
			\hline
			$(j_1, j_2)$ & $(\vec{n},\lambda)$ & $\textrm{CS}[\rho]$ & $\textrm{Tor}[\rho]$ & $\CO_{(a,b)}$ & $\textrm{CS}[\rho_{\alpha=0}]-\textrm{CS}[\rho]$
			\\
			\hline
			$(0, 0)$ & $(\frac{1}2 , \frac{1}2, \frac{1}2 , \frac{1}2)$ & $\frac{7}{48}$ & $2$ & $\CO_{(2,1)}=\CO_{(1,3)}$ & $\frac{1}2$
			\\
			\hline
			$(0, 2)$ & $(\frac{1}2,\frac{3}2, \frac{1}2 , \frac{1}2)$ & $\frac{31}{48}$ & $2$ & $\CO_{(2,3)} =\CO_{(1,1)}$ & $0$ 
			\\
			\hline
			$(1, 1)$ & $(1, 1, 1, 0)$ & $\frac{7}{12}$ & $1$ & $\CO_{(2,2)} = \CO_{(1,2)}$ & $\frac{1}{16}$
			\\
			\hline
		\end{tabular}
		\caption{Ising CFT  from $S^2 ((3,1),(4,3),(3,1))$. We use the characters-to-primaries map in \eqref{characters-to-primaries}. The result is compatible with that $S_{(1,1),(1,1)} = S_{(1,1),(2,1)} = \frac{1}2, S_{(1,1),(2,2)} = \frac{1}{\sqrt{2}} $ and $h_{(2,1)}= \frac{1}2, h_{(2,2)}= \frac{1}{16}$. } 
	\end{center}
\end{table}
\paragraph{Example: $M(5,2) =(\textrm{Lee-Yang CFT})$ from $ S^2 ((5,3),(2,1),(3,1))$}  We choose $(R,S) = (2,1)$.  There are  2 irreducible $SL(2,\mathbb{C})$ characters $\rho_{\vec{j}}$:
\begin{table}[h]
	\begin{center}
		\begin{tabular}{|c|c|c|c|c|c|} 
			\hline
			$(j_1, j_2)$ & $(\vec{n},\lambda)$ & $\textrm{CS}[\rho]$ & $\textrm{Tor}[\rho]$ & $\CO_{(a,b)}$ & $\textrm{CS}[\rho_{\alpha=0}]-\textrm{CS}[\rho]$
			\\
			\hline
			$(0, 0)$ & $(\frac{1}2, \frac{1}2, \frac{1}2, \frac{1}2)$ & $\frac{53}{120}$ & $\frac{1}{4} \left(5-\sqrt{5}\right)$ & $\CO_{(4,1)}=\CO_{(1,1)}$ & $0$
			\\
			\hline
			$(2, 0)$ & $(\frac{3}2, \frac{1}2, \frac{1}2, \frac{1}2)$ & $\frac{77}{120}$ & $\frac{1}{4} \left(5+\sqrt{5}\right)$ & $\CO_{(2,1)} =\CO_{(3,1)}$ & $\frac{4}5$ 
			\\
			\hline
		\end{tabular}
		\caption{Lee-Yang CFT  from $S^2 ((5,3),(2,1),(3,1))$.  The result is compatible with that $S_{(1,1),(1,1)} =  \sqrt{\frac{2}{5-\sqrt{5}}}, S_{(1,1),(3,1)} =  \sqrt{\frac{2}{5+\sqrt{5}}}$ and $h_{(3,1)}=-\frac{1}5$. } 
	\end{center}
\end{table}

\paragraph{Example: $M(5,4) =(\textrm{Tricritical Ising model})$ from $ S^2 ((5,-1),(4,5),(3,1))$}  We choose $(R,S) = (6,5)$.  There are  6 irreducible $SL(2,\mathbb{C})$ characters $\rho_{\vec{j}}$:
\begin{table}[h]
	\begin{center}
		\begin{tabular}{|c|c|c|c|c|c|} 
			\hline
			$(j_1, j_2)$ & $(\vec{n},\lambda)$ & $\textrm{CS}[\rho]$ & $\textrm{Tor}[\rho]$ & $\CO_{(a,b)}$ & $\textrm{CS}[\rho_{\alpha=0}]-\textrm{CS}[\rho_{\a}]$
			\\
			\hline
			$(0, 0)$ & $(\frac{1}2, \frac{1}2, \frac{1}2, \frac{1}2)$ & $\frac{37}{240}$ & $\frac{5}{2 \left(\frac{5}{8}-\frac{\sqrt{5}}{8}\right)}$ & $\CO_{(4,1)}=\CO_{(1,3)}$ & $\frac{1}2$
			\\
			\hline
			$(2, 0)$ & $(\frac{3}2, \frac{1}2, \frac{1}2, \frac{1}2)$ & $\frac{133}{240}$ & $\frac{5}{2 \left(\frac{5}{8}+\frac{\sqrt{5}}{8}\right)}$ & $\CO_{(2,1)}=\CO_{(3,3)}$ & $\frac{1}{10}$
			\\
			\hline
			$(0, 2)$ & $(\frac{1}2, \frac{3}2, \frac{1}2, \frac{1}2)$ & $\frac{157}{240}$ & $\frac{5}{2 \left(\frac{5}{8}-\frac{\sqrt{5}}{8}\right)}$ & $\CO_{(4,3)}=\CO_{(1,1)}$ & $0$
			\\
			\hline
			$(2, 2)$ & $(\frac{3}2, \frac{3}2, \frac{1}2, \frac{1}2)$ & $\frac{13}{240}$ & $\frac{5}{2 \left(\frac{5}{8}+\frac{\sqrt{5}}{8}\right)}$ & $\CO_{(2,3)}=\CO_{(3,1)}$ & $\frac{3}{5}$
			\\
			\hline
			$(1, 1)$ & $(1, 1, 1, 0)$ & $\frac{37}{60}$ & $\frac{5}{4 \left(\frac{5}{8}+\frac{\sqrt{5}}{8}\right)}$ & $\CO_{(2,2)}=\CO_{(3,2)}$ & $\frac{3}{80}$
			\\
			\hline
			$(3, 1)$ & $(2, 1, 1, 0)$ & $\frac{13}{60}$ & $\frac{5}{4 \left(\frac{5}{8}-\frac{\sqrt{5}}{8}\right)}$ & $\CO_{(4,2)}=\CO_{(1,2)}$ & $\frac{7}{16}$
			\\
			\hline
		\end{tabular}
		\caption{Tricritical Ising model  from $S^2 ((5,-1),(4,5),(3,1))$.  The result is compatible with the modular data of the tricritical Ising model, which can be evaluated from \eqref{conformal dimensions 1} and \eqref{modular S-matrix 2}}
	\end{center}
\end{table}
%One non-trivial evidence for the proposal is that the non-unitary TQFTs associated to the non-hyperbolic 3-manifolds have the same modular data as the corresponding minimal models. In \cite{Cho:2020ljj}, they study (2+1)d topological field theories, say  ${\rm TFT}[M]$, associated to non-hyperbolic 3-manifolds $M$ with certain properties. The modular data, $S$ and $T$ matrices, of the ${\rm TFT}[M]$ can be read off using the dictionaries in Table \ref{table : dictionaries}. The set of $SL(2,\mathbb{C})$ flat connections on the Seifert spaces and their invariants,  adjoint Reidemeister torsions and Chern-Simons invariants,  are explicitly given in \cite{Cui}. 
%
\\
\\
\\
\\
\section{Field theory description of $\CT_{(P,Q)}$}  \label{sec : field theory for T(P,Q)}

We now present a concrete and unified field-theoretic description for the $T_{\rm irred}[S^2 (\vec{p},\vec{q})]$ theory. By specializing the values of $((p_1, q_1),(p_2, q_2),(p_3,q_3))$ to  $((P,P-R), (Q,S),(3,1))$, the theory  becomes the $\CT_{(P,Q)}$ theory as proposed in \eqref{T(PQ) from SFS}. See \eqref{field  theory for T(P,Q)} for  the  proposed $\CT_{(P,Q)}$. 

In principle, the theory can be constructed using the general algorithm proposed in \cite{Dimofte:2011ju,Gang:2018wek}, which is based on a Dehn surgery representation of the 3-manifold using a hyperbolic link and an ideal triangulation of the link complement. As seen from various examples explored in \cite{Choi:2022dju}, however, one needs to consider different hyperbolic links for each $T_{\rm irred}[S^2 (\vec{p},\vec{q})]$. This makes it difficult to succinctly describe the field theories for all $S^2(\vec{p},\vec{q})$ in a unified manner.

\subsection{Field theory description of $T_{\rm irred}[S^2 (\vec{p},\vec{q})]$}
The Seifert fiber space $S^2 (\vec{p},\vec{q})$ depicted in Figure \ref{fig:SFS} can be alternatively represented as follows:
\begin{align}
\begin{split}
&S^2 (\vec{p},\vec{q}) = \left( (\Sigma_{0,3}\times S^1) \bigcup_{i=1}^3   (D_2 \times S^1)_i\right)/\sim
\\
&\textrm{ with } p_i A_i +q_i B_i \sim [S^1] \subset H_1 [\partial D_2, \mathbb{Z}]\;, \quad i=1,2,3\;.
\end{split}
\end{align}
Here, $\Sigma_{g=0,h=3}$ denotes a three-punctured sphere and $(A_i, B_i)$ are basis elements of $H_1 (\partial (\Sigma_{0,3}\times S^1),\mathbb{Z}_m)$  where $A_i$ (resp. $B_i$) represents the 1-cycle circling the $i$-th puncture (resp. the 1-cycle along the $S^1\in \partial (\Sigma_{0,3})\times S^1$). The equivalence relation $p_i A_i +q_i B_i \simeq [S^1] \in H_1[\partial D_2, \mathbb{Z}]$ corresponds to  the relation $x_i^{p_i} h^{q_i}=1$ in the fundamental group \eqref{fund group} and the relation $x_1 x_2 x_3 = 1$ comes from the same relation in $\pi_1 (\Sigma_{0,3})$.

\begin{figure}[h]
	\centering
	\includegraphics[width=0.45\textwidth,angle=0]{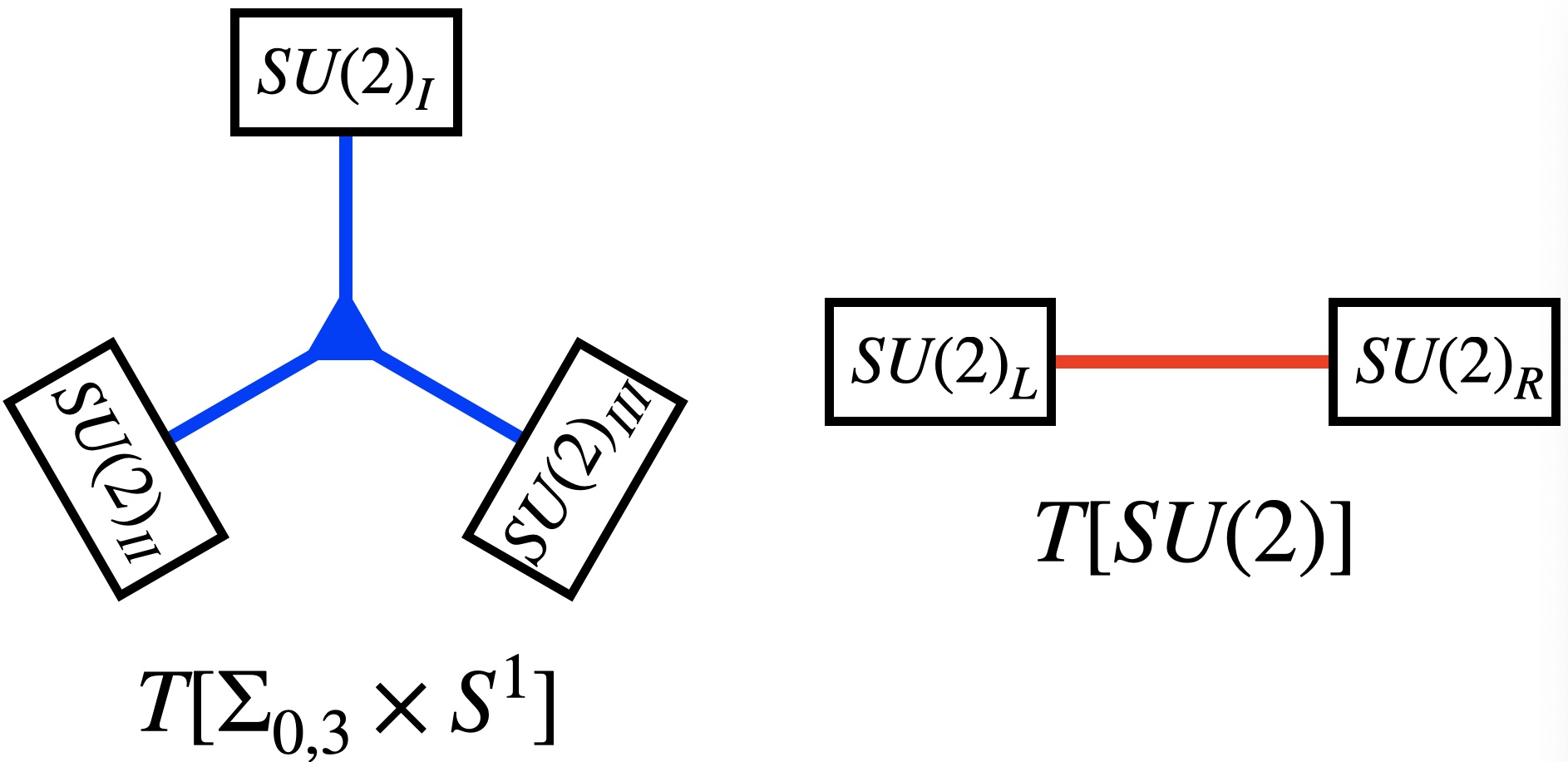}
	\includegraphics[width=0.45\textwidth,angle =0]{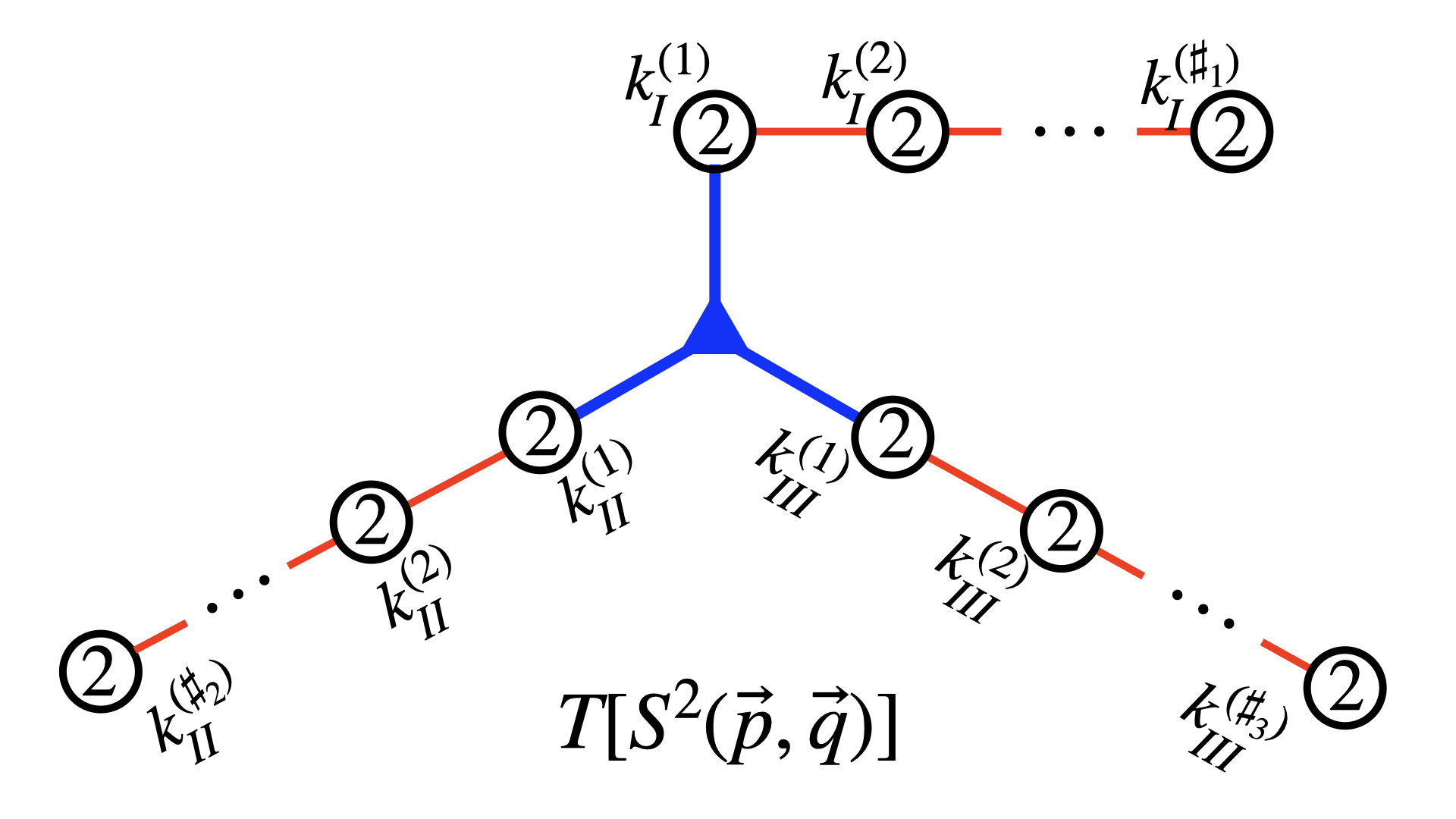}
	\caption{Generalized quiver diagrams for $T[\Sigma_{0,3}\times S^1]$, $T[SU(2)]$ and $T[S^2 (\vec{p},\vec{q})]$. The difference between $T_{\rm full}[S^2 (\vec{p},\vec{q})]$ and $T_{\rm irred}[S^2 (\vec{p},\vec{q})]$ arises only from different choices of the $T[\Sigma_{0,3}\times S^1]$ theory, either $T_{\rm full}[\Sigma_{0,3}\times S^1]$ or $T_{\rm irred}[\Sigma_{0,3}\times S^1]$. }
	\label{fig:quiver-SFS}
\end{figure}

Using the geometrical representation, the field theory $T[S^2 (\vec{p},\vec{q})]$ can be constructed as follows, see Figure \ref{fig:quiver-SFS}. 
First, we prepare the $T[\Sigma_{0,3}\times S^1]$ theory which is expected to be a $\CN=4$ theory with three $SU(2)$ flavor symmetries associated with the 3 boundary tori (punctures$\times S^1$). The theory is depicted by a blue trivalent vertex with three legs, and the boxes attached to the legs represent the three $SU(2)$ flavor symmetries.  Gluing the solid torus is a Dehn filling procedure, whose corresponding field-theoretic operation in 3D-3D correspondence has been studied in literature \cite{Pei:2015jsa,Alday:2017yxk,Gukov:2017kmk,Gang:2018wek,Assel:2022row}. In the Dehn filling operation, the $T[SU(2)]$ theory \cite{Gaiotto:2008ak}  plays an important role. The $T[SU(2)]$ is a $\CN=4$ theory with $SU(2)_L \times SU(2)_R$ flavor symmetry, (see Appendix \ref{appendix : Z[D(p,q)]} for details)  and is depicted by a red line with two boxes representing the two $SU(2)$ symmetries. The Dehn filling operation, $p_i A_i +q_i B_i \sim [S_1 ]\in H_1 [\partial D_2, \mathbb{Z}]$, at each $i$ corresponds to coupling the $T[\Sigma_{0,3}\times S^1]$ theory to $(\sharp-1)$-copies of $T[SU(2)]$ theories using the $i$-th $SU(2)$ flavor symmetry in $T[\Sigma_{0,3}\times S^1]$  and $2(\sharp-1)$ $SU(2)$s in the $T[SU(2)]^{\otimes (\sharp-1)}$, as described in 3rd quiver diagram in Figure \ref{fig:quiver-SFS}. The circle denotes the $\CN=3$ gauging the diagonal $SU(2)$, and the integer $k$ next to the circle denotes the Chern-Simons (CS) level. The number of $T[SU(2)]$s, $\sharp-1$, and the CS levels $\vec{k} = (k^{(1)}, \ldots, k^{(\sharp)})$ are related to  the Dehn filling slope $(p,q)$ as follows:
\begin{align}
\begin{split}
&\frac{q}p = \frac{1}{k^{(1)} - \frac{1}{k^{(2)} - \frac{1}{k^{(3)}- \ldots \frac{1}{k^{(\sharp)}}}}}\;.
\\
&\textrm{or equivalently,}
\\
& g(\vec{k} ) := T^{k^{(1)}} S T^{k^{(2)}} \ldots S T^{k^{(\sharp)}} = \pm \begin{pmatrix} q & * \\ p & * \end{pmatrix}\;.
\end{split} \label{p/q and k}
\end{align}
Here $S$ and $T \in SL(2,\mathbb{Z})$ are chosen as
\begin{align}
S= \begin{pmatrix}  0 & -1 \\ 1 & 0\end{pmatrix}, \quad T= \begin{pmatrix}  1 & 0 \\ 1 & 1\end{pmatrix}\;.
\end{align}

Especially when $p/q$ is an integer, i.e., $q=1$, the Dehn filling operation corresponds to the gauging of $SU(2)$ flavor symmetry with CS level $p$.
\paragraph{Field theory for $T_{\rm irred}\left[S^2((p_1,q_1),(p_2,q_2), (p_3,q_3))\right]$}
 Using the  prescription above, the field theory $T_{\rm full}[S^2 (\vec{p},\vec{q})]$ is constructed  as follows  \cite{Benini:2010uu,Eckhard:2019jgg,Assel:2022row}
\begin{align}
\begin{split}
&T_{\rm full} [\Sigma_{0,3}\times S^1] = (\textrm{$S^1$-reduction of 4D $T_{N=2}$ theory}) 
\\
&= (\textrm{a free theory with 8 half-hypermultiplets in $\mathbf{2}\otimes \mathbf{2}\otimes \mathbf{2}$ under the $SU(2)^3$} )\;.
\end{split}
\end{align}
The $T_{\rm irred}[S^2 (\vec{p},\vec{q})]$ can be constructed in the same way except that the $T_{\rm full}[\Sigma_{0,3}\times S^1]$ should be replaced by $T_{\rm irred}[\Sigma_{0,3}\times S^1]$.
 We propose that
\begin{align}
T_{\rm irred}\!\left[\Sigma_{0,3}\times S^{1}\right]
\end{align}
flows in the infrared to a topological field theory. This proposal is motivated by a general pattern observed in
$T_{\rm irred}[M]$ theories associated with non-hyperbolic
three-manifolds $M$ with torus boundaries: such theories typically
develop a mass gap and flow to a  unitary topological quantum field
theory in the infrared. Since $\Sigma_{0,3}\times S^{1}$ is
non-hyperbolic, it is natural to expect the same behavior in the present
case. At present, however, we do not know which specific TQFT arises as the
infrared fixed point of
$T_{\rm irred}\!\left[\Sigma_{0,3}\times S^{1}\right]$. 
 As we will see below, this proposal passes several non-trivial consistency checks.  
 
\begin{figure}[h]
	\centering
	\includegraphics[width=0.3\textwidth,angle =0]{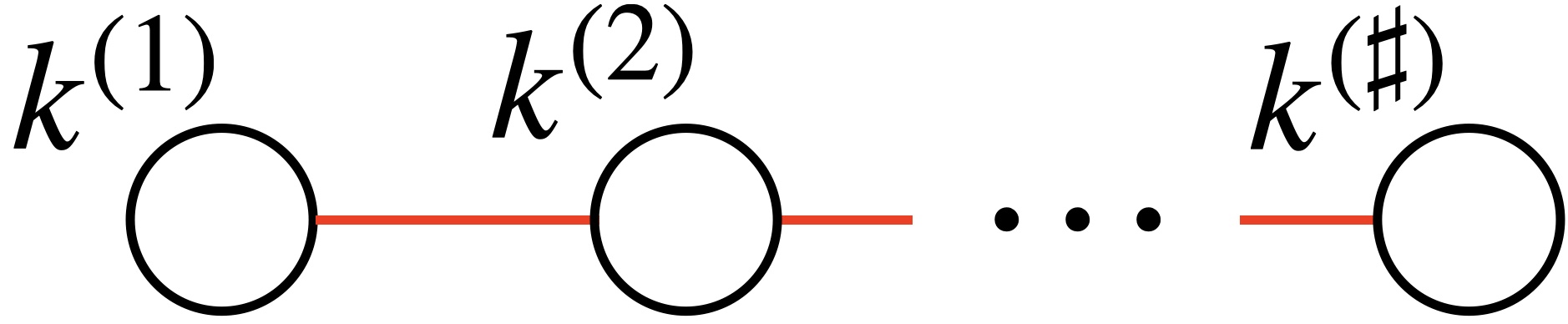}
	\caption{A quiver diagram for $D(\vec{k})$ theory.  }
	\label{fig:quiver}
\end{figure}

 Combining the above proposal with  the general Dehn surgery prescription in 3D-3D correspondence, we propose that
\begin{align}
\begin{split}
&T_{\rm irred}\left[M=S^2((p_1,q_1),(p_2,q_2), (p_3,q_3))\right]  
\\
&\sim  [ (\CD(p_1, q_1) \otimes \CD(p_2, q_2) \otimes \CD(p_3, q_3))]\;.
 \label{field theory for T[SFS]}
\end{split}
\end{align}
Here $\sim $ denotes the  IR equivalence in \eqref{two equivalence relations}.  The  theory $\CD(p,q)$ is defined as follows:
 \begin{align}
 \begin{split}
&D (\vec{k} ) \simeq  \CD(p,q) \otimes \textrm{TFT}[\vec{k} ] , \textrm{ where }
\\
&D(\vec{k} ):= \begin{cases} \frac{ T[SU(2)]^{\otimes (\sharp-1)} }{SU(2)^{(1)}_{k^{(1)}} \times SU(2)^{(2)}_{k^{(2)}}\ldots \times SU(2)^{(\sharp)}_{k^{(\sharp)}}},  &  \sharp \geq 2 
	\\
	\mathcal{N}=2\; \textrm{ pure }SU(2)_{k^{(1)}} \textrm{ CS theory}, & \sharp = 1
	\end{cases}
\end{split} \label{D(k) and D(p,q)}
 \end{align}
 Here $/G_k$ denotes $\CN=3$ gauging of $G$ symmetry with Chern-Simons level $k$.  The CS levels $\vec{k}  = (k^{(1)},\ldots, k^{(\sharp)})$  are related to the $(p,q)$ as in \eqref{p/q and k}.
The gauged $SU(2)$ symmetries are
\begin{align}
\begin{split}
&SU(2)^{(1)} \;:\; SU(2)_L^{(1)}:=SU(2)_L  \textrm{ of  the  1st } T[SU(2)]\;,
\\
&SU(2)^{(2\leq I \leq \sharp-1)} \;:\; \textrm{diagonal subgroup of } (SU(2)^{(I-1)}_R  \times SU(2)^{(I)}_L)\;,
\\
&SU(2)^{(\sharp)} \;:\; SU(2)_R^{(\sharp-1)}\;. 
\end{split}
\end{align} 
The field theory description can be summarized in the quiver diagram in Figure \ref{fig:quiver}.
The $D(\vec{k})$ theory has $(\mathbb{Z}_2)^{\otimes \sharp}$ 1-form symmetry originating from the center subgroup of  $SU(2)^{\sharp}$ gauge symmetry. The symmetry has non-trivial   `t Hooft anomalies \cite{Gang:2018wek} which can be characterized by  the following   action of the 4D anomaly theory
\begin{align}
S_{\rm anoamly}  = \pi  \int_{\CM_4}  \left( \sum_{I=1}^{\sharp} k^{(I)}\frac{\CP (\omega_2^{(I)})}2  +\sum_{J=1}^{\sharp-1}\omega_2^{(J)} \cup \omega_2^{(J+1)}  \right)\;.
\label{tHooft anomaly}\end{align} 
Here $\omega_2^{(I)} \in H^2 (\CM_4,\mathbb{Z}_2)$ is the two-form background field for each $\mathbb{Z}_2$ 1-form symmetry from $I$-th $SU(2)$ gauge symmetry and $\CP$  is the Pontryagin square operation. 
 By matching the anomaly with that of the decoupled topological field theory \cite{Hsin:2018vcg}, we expect  that the decoupled topological theory is given in the following form:
\begin{align}
\begin{split}
&\textrm{TFT}[\vec{k} = (k^{(1)}, k^{(2)},\ldots, k^{(\sharp)})] 
\\
&= \textrm{$U(1)^{\sharp}_\CK$  theory with mixed CS level } \CK_{IJ} =2\times  \begin{cases}
	+1 \textrm{ or }-1\;,  & |I-J|=1
	\\
     0\;, & I=J  \textrm{ and } k^{(I)} \in 2\mathbb{Z}
     \\
     +1 \textrm{ or }-1 \;, & I=J  \textrm{ and } k^{(I)} \in 2\mathbb{Z}+1
	\\
	0\;,   & |I-J|>1
\end{cases} 
\end{split}\;. \label{decoupled TQFT}
\end{align}
Refer to \cite{Comi:2023lfm} for a similar analysis done for $T_{\rm full }[S^2 ((k_1,1),(k_2,1),(k_3,1))]$ theory. 
See also Appendix \ref{appendix: decoupled TQFT} for more details of the decoupled TQFT.  
By removing the decoupled TQFT from $D(\vec{k} )$, we obtain the $\CD(p,q)$. We expect the following properties of the $\CD(p,q)$ theory
\begin{align}
\begin{split}
&i) \;\CD(p,q) \textrm{ does not depend on the choice of $\vec{k} $ for given $(p,q)$.  }\; ,
\\
& ii) \;\CD(p,q) \simeq \CD(p, q+ p \mathbb{Z})\;. \label{Two properties of D(p,q)}
\end{split}
\end{align}
\begin{figure}[h]
	\centering
	\includegraphics[width=0.6\textwidth,angle =0]{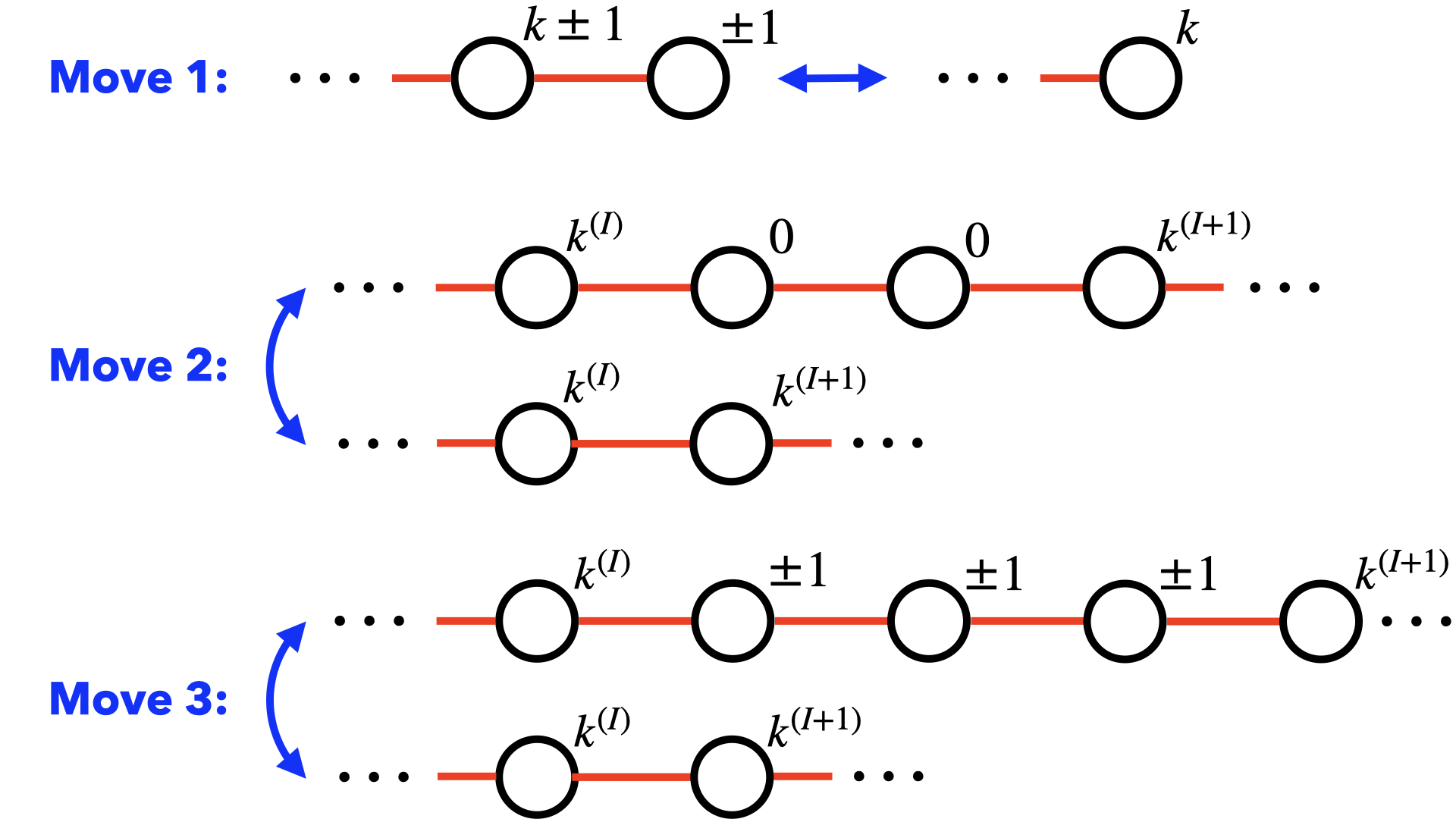}
	\caption{Three basic dualities (modulo topological sectors) guaranteeing the independence of $D(p,q)$ on the choices of $\vec{k}$.  The 2nd and 3rd moves follow from the $SL(2,\mathbb{Z})$ structure of the duality wall theory. The 1st move comes from the IR duality between $(T[SU(2)]/(SU(2)_R)_{\pm 1} )\sim (\textrm{empty theory only with background CS level $\mp 1$ for }SU(2)_L)$. See the computation in \eqref{SCI for TSU2/SU(2)} for a non-trivial check of the duality using the superconformal index. The same moves were used in constructing $T_{\rm full}[M]$ for 3-manifolds $M$ associated to plumbed graphs in \cite{Gukov:2017kmk}. }
	\label{fig:Kirby-moves}
\end{figure}
The first property follows from the basic IR dualities (modulo topological sectors) depicted in Figure \ref{fig:Kirby-moves}, which imply that
\begin{align}
D(\vec{k}_1) \sim D(\vec{k}_2) \quad  \textrm{ if $g(\vec{k}_1) =g(\vec{k}_2) \cdot  T^{\pm 1} S  T^{\pm 1} $} \;.
\end{align}
Note that $g(\vec{k}) = \pm \left(
\begin{array}{cc}
q & * \\
p & * \\
\end{array}
\right) $ and the right multiplication of $T^{\pm 1 }ST^{\pm 1} =\left(
\begin{array}{cc}
\mp 1 & -1 \\
0 & \mp 1 \\
\end{array}
\right)$ does not change the slope $(p,q)$. 
 The 2nd property follows from the move 1 in the figure with reversed left/right orientation on each quivers, which implies that  
\begin{align}
D(\vec{k}_1) \sim D(\vec{k}_2) \quad\textrm{ if $g(\vec{k}_1) =  T^{\pm 1} S T^{\pm 1}  \cdot g(\vec{k}_2)$}\;.
\end{align}  
The left multiplication of $T^{\pm 1 }ST^{\pm 1}$ changes the $(p,q)$ to  $(p, q\pm  p)$. The above IR equivalence $\sim$ modulo topological sector can be promoted into the  IR equivalence $\simeq$ in \eqref{two equivalence relations} if the decoupled $\textrm{TFT}(\vec{k})$s on both sides are removed.

From the two properties in \eqref{Two properties of D(p,q)}, it is easy to see that (for $|p|\geq 2$):
\begin{align}
\begin{split}
&\textrm{If  $q = \pm 1 \;(\textrm{mod }p)$,}
\\
 &\textrm{the $\CD(p,q)$ has a mass gap and flows to a unitary TQFT in the IR} \label{gapped case of D(p,q)}
 \end{split}
\end{align}
since 
\begin{align}
\begin{split}
 &\CD(p,\pm 1 +p \mathbb{Z}) \simeq  \CD(p, \pm 1) \sim D(\vec{k} = (\pm p)) =(\textrm{$\CN=3$  $SU(2)_{\pm p}$})  
 \\
 & \simeq  (\textrm{$\CN=2$  $SU(2)_{\pm p}$})  \simeq SU(2)_{ \pm p - 2 \times \textrm{sign}(\pm p)} \;. \nonumber
 \end{split}
 \end{align}
 The pure $\CN=3$  CS theory with non-zero CS level is IR equivalent to pure $\CN=2$ CS theory since the adjoint chiral multiplet in the $\CN=3$ multiplet has a superpotential mass term and can be integrated out. 
The pure $\CN=2$  Chern-Simons theory $SU(2)_k$ contains an auxiliary massive gaugino, and integrating it out induces a CS level shift by $-2 \times  \textrm{sign}(k) $. Furthermore, one can check that
\begin{align}
\CD(2,1 + 2\mathbb{Z}) \simeq \CD(3,1+3\mathbb{Z})\simeq (\textrm{a trivial theory}), \label{triviality of D(2,1) and D(3,1)}
\end{align}
The first follows from the fact that $SU(2)_{2-2}=SU(2)_0$ is a trivial theory, while the 2nd follows from that  $SU(2)_{3-2}=SU(2)_1 \simeq U(1)_2$ and $\textrm{TFT}[\vec{k}= (3)]$ in \eqref{decoupled TQFT} is again  $U(1)_{2}$.
On the other hand, we expect that 
\begin{align}
\begin{split}
&\textrm{If  $q \neq  \pm 1 \;(\textrm{mod }p)$,}
\\
&\textrm{the $\CD(p,q)$  flows to an $\CN=4$ rank-0 SCFT in the IR} \label{N=4 rank-0 case of D(p,q)}.
\end{split}
\end{align}
 The non-zero CS terms lift all the Coulomb/Higgs branches of $T[SU(2)]$ theories. The UV $\CN=3$ will be enhanced to $\CN=4$ in the IR thanks to the nilpotency properties of moment maps of the $T[SU(2)]$ theory \cite{Gaiotto:2008ak,Gang:2018huc,Garozzo:2019ejm}. 
Along with the proposal in  \eqref{field theory for T[SFS]}, \eqref{gapped case of D(p,q)}  and \eqref{N=4 rank-0 case of D(p,q)} are compatible with the expected IR phases of $T[S^2 (\vec{p},\vec{q})]$ given in \eqref{IR phases of T[SFS]}.

\paragraph{Comparison with $T_{\rm irred}[S^2 ((\vec{p},\vec{q}))]$s in  \cite{Choi:2022dju}} Based on a  Dehn surgery representation  of $S^2 ((\vec{p},\vec{q}))$ with hyperbolic knots,  the field theories of $T_{\rm irred}[S^2 ((\vec{p},\vec{q}))]$ for various $(\vec{p},\vec{q})$s were analyzed in \cite{Choi:2022dju}. For instance, the 3D theory $T_{\rm irred}[M]$ for $M=(S^3\backslash \mathbf{5}_2)_{p}$ (the 3-manifold obtained by a Dehn surgery on $\mathbf{5}_2$ knot, a.k.a the 3-twist knot, with an integral slope $p$) is given by 
\begin{align}
\begin{split}
&T_{\rm irred}[(S^3\backslash \mathbf{5}_2)_{p}] 
\\
&= \begin{cases}
	\frac{\left( U(1)_{-\frac{1}2}\times SU(2)_{p-2}  \textrm{ $\CN=2$ gauge theory coupled to a chiral in (Adj)}_{+1} \right)}{\mathbb{Z}_2}\;, & p \in 2\mathbb{Z}
	\\
	\frac{(U(1)_{-\frac{1}2}\times SU(2)_{p-2}  \textrm{ $\CN=2$ gauge theory coupled to a chiral in (Adj)}_{+1})\otimes U(1)_{\pm 2}}{\mathbb{Z}^{\rm diag}_2}\;, & p \in 2\mathbb{Z}+1
\end{cases}
\end{split}
\end{align}
The theory in the numerator has a $\mathbb{Z}_2$ 1-form symmetry originating from the center $\mathbb{Z}_2$ subgroup of the $SU(2)$ gauge group. For even $p$, the $\mathbb{Z}_2$ 1-form symmetry is non-anomalous (i.e., having a trivial `t Hooft anomaly) and thus can be gauged. For odd $p$,  however, the 1-form symmetry has a non-trivial `t Hooft anomaly, requiring it to be tensored with the $U(1)_{\pm 2}$ theory, which has the same anomalous 1-form symmetry, before gauging the non-anomalous diagonal $\mathbb{Z}_2$ 1-form symmetry.

Topologically it is known that \cite{dunfield:2018census}
\begin{align}
(S^3\backslash \mathbf{5}_2)_{p} = \begin{cases}
	S^2 ((2,1),(3,1),(11,-9))\;, & p=1
	\\
	S^2 ((2,1),(4,1),(7,-5))\;, & p=2
	\\
	S^2 ((3,1),(3,1),(5,-3))\;, & p=3
\end{cases}
\end{align}
According to  our proposal in \eqref{field theory for T[SFS]}, we expect the following dual description\footnote{Using $\frac{-9}{11} = \frac{1}{-1-\frac{1}{4-\frac{1}{-2}}}$, $\frac{-5}{7} = \frac{1}{-1-\frac{1}{2-\frac{1}{-2}}}$ and $-\frac{3}5 = \frac{1}{-2-\frac{1}{-3}}$.}
\begin{align}
T_{\rm irred}[(S^3\backslash \mathbf{5}_2)_{p} ]  \sim   \begin{cases}
D(\vec{k}= (-1,4,-2))\;, & p=1
\\
D(\vec{k}= (-1,2,-2))\;, & p=2
\\
D(\vec{k}= (-2,-3)) \;, & p=3
\end{cases}
\end{align}
Note that $\CD(p,1)$ with $p\in \mathbb{Z}$ is a (unitary) TQFT and thus ignored in the above. Using the expression in \eqref{SCI for D(k)}, one can compute the superconformal indices:
\begin{align}
\begin{split}
&\CI^{\rm sci}_{D(\vec{k})} (\eta, \nu=0)  \;\;\textrm{    with }\vec{k}= (-1,4,-2)
\\
&=  1- q- \left( \eta +\frac{1}\eta \right) q^{3/2} - 2 q^2 -\eta^{-1} q^{5/2}+(\eta^2-1) q^3 +\left(\eta -\frac{1}\eta\right)q^{7/2} +0 q^4 +\ldots
\\
&\CI^{\rm sci}_{D(\vec{k})} (\eta, \nu=0)   \;\;\textrm{    with }\vec{k}= (-1,2,-2)
\\
&=  1- q- \left( \eta +\frac{1}\eta \right) q^{3/2} - 2 q^2 -\eta^{-1} q^{5/2}+(\eta^2-1) q^3 +\left(\eta -\frac{1}\eta\right)q^{7/2} + \eta^2 q^4 +\ldots
\\
&\CI^{\rm sci}_{D(\vec{k})}  (\eta, \nu=0) \;\; \textrm{    with }\vec{k}= (-2,-3)
\\
&=  1- q- \left( \eta +\frac{1}\eta \right) q^{3/2} - 2 q^2 - \left( \eta +\frac{1}\eta \right) q^{5/2}- 2 q^3 - \left(\eta +\frac{1}\eta \right)q^{7/2} - 2 q^4 +\ldots\;.
\end{split}
\end{align}
These results match the superconformal indices  given in \cite{Choi:2022dju} for the $T_{\rm irred}[(S^3\backslash \mathbf{5}_2)_{p}]$ theory with $p=1,2,3$ respectively. For rank-0 SCFTs, the superconformal index is defined as: 
\begin{align}
\CI^{\rm sci} (\eta, \nu) :=\textrm{Tr} (-1)^{R_\nu} q^{\frac{ R_\nu}{2}+j_3} \eta^A\;, \label{SCI}
\end{align}
where $A$ and $R_\nu$ are defined in \eqref{charge A} and \eqref{R-symmetry mixing}.
\subsection{Field theory for $\CT_{(P,Q)}$}
Combining \eqref{T(PQ) from SFS} and \eqref{field theory for T[SFS]}, we arrive at the final field-theory description:
\begin{align}
\begin{split}
\CT_{(P,Q)}
\;&\simeq\;
T_{\rm irred}\!\left[S^2\bigl((P, P\!-\!R),(Q,S),(3,1)\bigr)\right]
\\
&\simeq\;
\begin{cases}
\CD(P,P\!-\!R)\otimes \CD(Q,S)\otimes U(1)_{\pm 2},
& P,Q\in 2\mathbb{Z}+1,
\\[4pt]
\bigl(\CD(P,P\!-\!R)\otimes \CD(Q,S)\otimes U(1)_{2}\otimes U(1)_{\pm 2}\bigr)\big/\mathbb{Z}_2,
& \text{otherwise}.
\end{cases}
\end{split}
\label{field  theory for T(P,Q)}
\end{align}
Here the theory $\CD(p,q)$ is defined in \eqref{D(k) and D(p,q)}.
\textbf{This is the main result of our paper.} The theory $\CD(P,P\!-\!R)$ depends on $R$ only modulo $P$, up to strong equivalence $\simeq$, while $\CD(Q,S)$ depends on $S$ only modulo $Q$; see ii) of \eqref{Two properties of D(p,q)}.
For a given coprime pair $(P,Q)$, the integers $(R,S)$ are uniquely determined, up to shifts by integer multiples of $(P,Q)$, by the Diophantine relation $PS-QR=1$.
Such shifts do not affect the corresponding theories beyond strong equivalence.
Consequently, the theories $\CD(P,P\!-\!R)$ and $\CD(Q,S)$ are uniquely determined by the choice of $(P,Q)$, up to strong equivalence.

The quotient $/\mathbb{Z}_2$ denotes gauging a particular $\mathbb{Z}_2$ one-form symmetry of the theory in the numerator.
Note that the numerator theory generally admits several $\mathbb{Z}_2$ one-form symmetries, and the quotient refers to a specific choice among them. Since the precise structure of the unitary TQFT $T[\Sigma_{0,3}\times S^1]$ is not known, 
we infer the required topological operations—namely the appearance of a decoupled $U(1)_{\pm2}$ factor and the gauging of a $\mathbb{Z}_2$ one-form symmetry—from indirect consistency conditions. In particular, in the first case we include a decoupled topological field theory $U(1)_2$ or $U(1)_{-2}$ so that
(recalling $\CZ_b^{\rm con}:=\CZ^{S^3_b}(M=0,\nu=0)$)
\begin{align}
\begin{split}
\bigl|\CZ^{\rm con}_{b=1}(\CT_{(P,Q)})\bigr|
&=
\Bigl|
\CZ^{\rm con}_{b=1}\!\left(\CD(P,P\!-\!R)\right)
\;
\CZ^{\rm con}_{b=1}\!\left(\CD(Q,S)\right)
\;
\times \frac{1}{\sqrt{2}}
\Bigr|
\\
&=
\sqrt{\frac{8}{PQ}}\,
\sin\!\left(\frac{\pi}{P}\right)
\sin\!\left(\frac{\pi}{Q}\right)
=
\min_{\alpha}\bigl|S_{0\alpha}\bigr|\;\text{of } M(P,Q)\; .
\end{split}
\end{align}
Here we have used the result in \eqref{S3 D(p,q)} together with the fact that the $S^3$ partition function of $U(1)_{\pm2}$ equals $1/\sqrt{2}$.
The decoupled $U(1)_{\pm2}$ theory possesses a $\mathbb{Z}_2$ one-form symmetry generated by an anyon (topological line defect) of topological spin $\pm \tfrac{1}{4}$ modulo $1$.
This is consistent with the existence of a primary operator of conformal dimension $\pm \tfrac{1}{4}$ modulo $1$ in the minimal model $M(P,Q)$ for $P,Q\in 2\mathbb{Z}+1$.

When $P$ (resp.~$Q$) is even, the theory $\CD(P,P-R)$ (resp.~$\CD(Q,S)$) has a non-anomalous $\mathbb{Z}_2$ 1-form symmetry generated by an anyon with topological spin $0$ or $1/2$ \cite{Hsin:2018vcg}, see \eqref{Z2 in even p}.  The theory in the numerator has three $\mathbb{Z}_2$ 1-form symmetries and $\mathbb{Z}_2$ denotes gauging a $\mathbb{Z}_2$ subgroup of the $(\mathbb{Z}_2)^{\otimes 3}$. The $\pm$ in the $ U(1)_2 \otimes U(1)_{\pm 2}$ is chosen such that the $\mathbb{Z}_2$ 1-form symmetry is bosonic (i.e., the symmetry generating anyon has a topological spin $0$).

When $Q=P+1$, the $(R,S)$ can be chosen as $ (-1,-1)$ and the $\CT_{(P,Q)}$ theory becomes
\begin{align}
\begin{split}
&(\CD(P,P+1)\otimes \CD (P+1,-1) \otimes U(1)_2 \otimes U(1)_{\pm 2})/\mathbb{Z}_2\;,
\\
& \simeq \frac{SU(2)_{(P-2)} \otimes SU(2)_{-(P-1)} \otimes U(1)_2}{\mathbb{Z}^{\rm diag}_2}\;.
\end{split}
\end{align}
Here the $\mathbb{Z}_2$ subgroup is chosen as the diagonal $\mathbb{Z}_2$ of $(\mathbb{Z}_2)^{\otimes 3}$. This is the coset description of unitary minimal model $\CT_{(P,P+1)}$. We use the fact that 
\begin{align} 
\begin{split}
& \textrm{For }P\in 2\mathbb{Z}_{\geq 1},\;  \CD (P+1, -1 ) \otimes  U(1)_{\pm 2}\simeq SU(2)_{-(P-1)}  \textrm{ and } \;\CD (P,P+1) \simeq SU(2)_{P-2}  \;,
\\
& \textrm{For }  P\in 2\mathbb{Z}_{\geq 1}+1,  \; \CD (P+1,-1 ) \simeq SU(2)_{-(P-1)}\textrm{ and }\;\CD (P,P+1)\otimes U(1)_{\mp 2} \simeq SU(2)_{(P-2)} \;.
\nonumber
\end{split}
\end{align}
The  2D  chiral central charge of the theory can be computed as follows using   that $c_{2d}(SU(2)_k) = \frac{3k}{|k|+2}$ and $c_{2d}(U(1)_2)=1$,
\begin{align}
c_{2d}=  \frac{3(P-2)}{P} -\frac{3 (P-1)}{P+1} +1 =1 - \frac{6}{P(P+1)}\;,
\end{align} 
which matches \eqref{central charge}.

\paragraph{Example : $(P,Q)=(3,5)$} Choosing $(R,S)=(1,2)$ and using equations \eqref{field  theory for T(P,Q)} and \eqref{triviality of D(2,1) and D(3,1)}, we find:
\begin{align}
\CT_{(P,Q)} \simeq  \CD(3,2)\otimes \CD(5,2) \otimes U(1)_{\pm 2} \simeq \CD(5,2)\otimes U(1)_{\pm 2}\;.
\end{align}
From that $\frac{2}5 = \frac{1}{3- \frac{1}{2}}$, we obtain
\begin{align}
\CD(5,2) \otimes {\rm TFT}[\vec{k}]\simeq  D(\vec{k})\textrm{ with }\vec{k}=(3,2)\;.
\end{align}
Utilizing the explicit computation from \eqref{H and F of D(k)}, we confirm that the set $\mathbf{HF}:={(\CH^{-1/2},\CF/\CF_{\alpha=0})}$ of the $D((3,2))$ theory in the $A$-twisting limit can be factorized as follows:
\begin{align}
\begin{split}
&\mathbf{HF}^{A} \textrm{ of } D((3,2))
\\
&=  \bigg{\{}\left(\frac{4}{\sqrt{15}} \sin (\frac{\pi}3 )\sin (\frac{2\pi}5), 1 \right) , \left(\frac{4}{\sqrt{15}} \sin (\frac{\pi}3 )\sin (\frac{\pi}5), e^{\frac{2\pi i }{5}} \right) \bigg{\}} \times \bigg{\{} \left(\frac{1}2, 1\right)^{\otimes 2}, \left(\frac{1}2, i\right), \left(\frac{1}2, \frac{1}i\right)\bigg{\}}\;.
\end{split}
\end{align}
The second factor can be interpreted as the contribution from the decoupled ${\rm TFT}[\vec{k}]$. By taking the product of the first factor, which is from $\CD(5,2)$, with the $\mathbf{HF}=\{ (|S_{0\alpha}|, e^{2\pi i h_\alpha})\}$ of $U(1)_{-2}$, we obtain the set for the $\CT_{(3,5)}$ theory:
\begin{align}
\begin{split}
&\mathbf{HF}^{A} \textrm{ of } \CT_{(3,5)}
\\
&=  \bigg{\{}\left(\frac{4}{\sqrt{15}} \sin (\frac{\pi}3 )\sin (\frac{2\pi}5), 1 \right) , \left(\frac{4}{\sqrt{15}} \sin (\frac{\pi}3 )\sin (\frac{\pi}5), e^{\frac{2\pi i }{5}} \right) \bigg{\}} \times \bigg{\{} \left(\frac{1}{\sqrt{2}}, 1\right),\left(\frac{1}{\sqrt{2}},\frac{1}i \right)\bigg{\}}\;.
\end{split}
\end{align}
This set nicely matches with the set of $(|S_{0\alpha}|, e^{2\pi i h_\alpha})$ for $M(3,5)$, as expected from the dictionaries in Table \ref{table : dictionaries}.

Alternatively, using $\frac{2}5= \frac{1}{2-\frac{1}{-2}}$, we have:
\begin{align}
\CD(5,-2) \otimes {\rm TFT}[\vec{k}]\simeq  D(\vec{k})\textrm{ with }\vec{k}=(2,-2)\;.
\end{align}
In this case, we find:
\begin{align}
\begin{split}
&\mathbf{HF}^{A} \textrm{ of } D((2,-2))
\\
&=  \bigg{\{}\left(\frac{4}{\sqrt{15}} \sin (\frac{\pi}3 )\sin (\frac{2\pi}5), 1 \right) , \left(\frac{4}{\sqrt{15}} \sin (\frac{\pi}3 )\sin (\frac{\pi}5), e^{\frac{2\pi i }{5}} \right) \bigg{\}} \times \bigg{\{} \left(\frac{1}2, 1\right)^{\otimes 3}, \left(\frac{1}2, -1\right) \bigg{\}}\;.
\end{split}
\end{align}
Again, the second factor can be interpreted as the contribution from the decoupled ${\rm TFT}[\vec{k}]$. By omitting the topological factor, we obtain the set for $\CD(5,2)$. It is equivalent to the previous set obtained using $D((3,2))$.

\paragraph{Example : $(P,Q)=(3,8)$}  Choosing $(R,S)=(1,3)$, we have:
\begin{align}
\CT_{(3,8)} \simeq \left(\CD(3,2)\otimes \CD(8,3)\otimes U(1)_2\otimes U(1)_{\pm 2}\right)/\mathbb{Z}_2 \simeq \CD(8,3)\otimes \frac{\left( U(1)_2 \otimes U(1)_{\pm 2}\right)}{\mathbb{Z}^{\rm diag}_2} \simeq \CD(8,3)   \;.
\end{align}
%\otimes U(1)_{\pm 2}
Using $\frac{3}8 = \frac{1}{3- \frac{1}{3}}$, we obtain:
\begin{align}
\CD(8,3) \otimes {\rm TFT}[\vec{k}]\simeq  D(\vec{k})\textrm{ with }\vec{k}=(3,3)\;.
\end{align}
Using the explicit computation  in \eqref{H and F of D(k)}, we  confirm that the set  $\mathbf{HF}:=\{(\CH^{-1/2},\CF/\CF_{\alpha=0})\}$ of $D((3,3))$ theory in the $A$-twisting limit can be factorized as follows:
\begin{align}
\begin{split}
\mathbf{HF}^{A} &\textrm{ of } D((3,3))
\\
=\bigg{\{}
&\left(\frac{1}{2} , e^{\frac{29 i \pi
	}{16}}  \right),
\left(\frac{1}{2 \sqrt{2}} , e^{\frac{25
		i \pi }{16}}  \right),
\left(\frac{1}{2 \sqrt{2}} , e^{\frac{25
		i \pi }{16}}  \right),
\left(\frac{1}{2} \sin \left(\frac{\pi
}{8}\right) , i  \right), 
\\
&\left(\frac{1}{2} \sin \left(\frac{\pi
}{8}\right) , \frac{1}{i}  \right),
\left(\frac{1}{2} \cos \left(\frac{\pi
}{8}\right) , 1  \right),
\left(\frac{1}{2} \cos \left(\frac{\pi
}{8}\right) , -1 \right)
\bigg{\}}
\times \bigg{\{} \left(\frac{1}{\sqrt{2}} , 1\right), \left(\frac{1}{\sqrt{2}} , i \right)\bigg{\}}\;.
\end{split}
\end{align}
The 2nd factor can be regarded as the contribution from the decoupled ${\rm TFT}[\vec{k}]$ and 1st factor is from $\CD(8,3)$. The first factor  nicely matches  the set of $(|S_{0\alpha}|, e^{2\pi i h_\alpha})$ of $M(3,8)$, as expected from the dictionaries in Table \ref{table : dictionaries}.

\paragraph{\boldmath Example : $(P,Q)=(5,8)$}  
We choose $(R,S)=(11,7)$, so that
\begin{align}
\begin{split}
\CT_{(5,8)} &\simeq  (\CD(5,7)\otimes \CD(8,-3)\otimes U(1)_2\otimes U(1)_{-2})/\mathbb{Z}_2 \\
&\simeq  \CD(5,2)\otimes (\CD(8,-3)\otimes U(1)_2\otimes U(1)_{-2})/\mathbb{Z}_2^{\text{diag}} \;.
\end{split}
\end{align}
%\otimes U(1)_{\pm 2}

Using $\frac{2}{5} = \frac{1}{3- \frac{1}{2}}$, we get
\begin{align}
\CD(5,2) \otimes {\rm TFT}[\vec{k}]\simeq  D(\vec{k})\textrm{ with }\vec{k}=(3,2)\;.
\end{align}
From \eqref{H and F of D(k)}, the set $\mathbf{HF}:=\{(\CH^{-1/2},\CF/\CF_{\alpha=0})\}$ for the $D((3,2))$ theory in the $A$-twisting limit is given by
\begin{align}
\begin{split}
\mathbf{HF}^{A} &\textrm{ of } D((3,2))
\\
=\bigg\{
  &\left( \frac{2}{\sqrt{10 - 2\sqrt{5}}},\ e^{\frac{i\pi}{10}} \right),\ 
  \left( \sqrt{\frac{2}{5 + \sqrt{5}}},\ i \right)
\bigg\} %\\
\times \bigg{\{} \left(\frac{1}2, 1\right)^{\otimes 2}, \left(\frac{1}2, i\right), \left(\frac{1}2, i^{-1}\right)\bigg{\}}\;.
\end{split}
\end{align}
The first factor corresponds to $\CD(5,2)$; the second to the decoupled ${\rm TFT}[\vec{k}]$. 

Using $\frac{-3}{8} = \frac{1}{-3- \frac{1}{-3}}$, we have
\begin{align}
\CD(8,-3) \otimes {\rm TFT}[\vec{k}]\simeq  D(\vec{k})\textrm{ with }\vec{k}=(-3,-3)\;.
\end{align}
From \eqref{H and F of D(k)}, one can obtain the set $\mathbf{HF}:=\{(\CH^{-1/2},\CF/\CF_{\alpha=0})\}$ for the $D((-3,-3))$ theory in the $A$-twisting limit:
\begin{align}
\begin{split}
\mathbf{HF}^{A} &\textrm{ of } D((-3,-3))
\\
=    \bigg\{
    &\left(\frac{1}{2\sqrt{2}},\, e^{-\frac{i\pi}{16}}\right),\;
    \left(\frac{1}{2\sqrt{2}},\, e^{-\frac{i\pi}{16}}\right),\;
    \left(\frac{1}{2},\, e^{-\frac{5 i\pi}{16}}\right),\\
    &\left(\frac{\csc\!\left(\frac{\pi}{8}\right)}{4\sqrt{2}},\, -1\right),\;
    \left(\frac{\csc\!\left(\frac{\pi}{8}\right)}{4\sqrt{2}},\, 1\right),\;
    \left(\frac{1}{2}\sin\!\left(\frac{\pi}{8}\right),\, -1\right),\;
    \left(\frac{1}{2}\sin\!\left(\frac{\pi}{8}\right),\, 1\right)
    \bigg\} \\
\times \bigg{\{} &\left(\frac{1}{\sqrt{2}}, 1\right), \left(\frac{1}{\sqrt{2}}, i\right) 
\bigg{\}}\;.
\end{split}
\end{align}
The first factor derives from $\CD(8,-3)$, while the second comes from the decoupled ${\rm TFT}[\vec{k}]$. 

The product of the first factor (from $\CD(5,2)$) and the second factor (from $(\CD(8,-3)\otimes U(1)_2\otimes U(1)_{-2})/\mathbb{Z}_2^{\text{diag}}$) %\footnote{see the footnote preceding \eqref{T(6,17)}} 
yields the following set for the $\CT_{(5,8)}$ theory:
\begin{align}
\begin{split}
\mathbf{HF}^{A} &\textrm{ of } \CT_{(5,8)}
\\
=  \bigg\{
 &\left(\frac{1}{2\sqrt{2}},\, e^{-\frac{i\pi}{16}}\right), \left(\frac{1}{2\sqrt{2}},\, e^{-\frac{i\pi}{16}}\right), \left(\frac{1}{2},\, e^{-\frac{5 i \pi}{16}}\right),\\
 &\left(\frac{\csc\!\left(\frac{\pi}{8}\right)}{4\sqrt{2}},\, -1\right), \left(\frac{\csc\!\left(\frac{\pi}{8}\right)}{4\sqrt{2}},\, 1\right), \left(\frac{1}{2}\sin\!\left(\frac{\pi}{8}\right),\, -i\right), \left(\frac{1}{2}\sin\!\left(\frac{\pi}{8}\right),\, i\right)
\bigg\} \\
\times 
\bigg\{
  &\left( \frac{2}{\sqrt{10 - 2\sqrt{5}}},\ e^{\frac{i\pi}{10}} \right),\ 
  \left( \sqrt{\frac{2}{5 + \sqrt{5}}},\ i \right)
\bigg\}%(=Lorentz fractionalization of \overline{SU(2)_4})
\;.
\end{split}
\end{align}
This agrees with the set $(|S_{0\alpha}|, e^{2\pi i h_\alpha})$ of $M(5,8)$ as in Table \ref{table : dictionaries}.

\paragraph{\boldmath Example : $(P,Q)=(7,11)$}  
We select $(R,S)=(14,9)$, so that
\begin{align}
\begin{split}
\CT_{(7,11)} \simeq  \CD(11,-3)\otimes \CD(7,9)\otimes U(1)_2 \simeq  \CD(11,-3)\otimes \CD(7,2)\otimes U(1)_2 \;.
\end{split}
\end{align}
%\otimes U(1)_{\pm 2}
Using $\frac{-3}{11} = \frac{1}{-4- \frac{1}{-3}}$, we obtain
\begin{align}
\CD(11,-3) \otimes {\rm TFT}[\vec{k}]\simeq  D(\vec{k})\textrm{ with }\vec{k}=(-4,-3)\;.
\end{align}
From \eqref{H and F of D(k)}, $\mathbf{HF}:=\{(\CH^{-1/2},\CF/\CF_{\alpha=0})\}$ for the $D((-4,-3))$ theory in the $A$-twisting limit reads 
\begin{align}
\begin{split}
\mathbf{HF}^{A} &\textrm{ of } D((-4,-3))
\\
=\bigg\{\, &\left(
\frac{2}{\sqrt{11}} \cos\!\left(\frac{\pi}{22}\right),
\, e^{\frac{8 i \pi}{11}}
\right),\;
\left(
\frac{2}{\sqrt{11}} \cos\!\left(\frac{3\pi}{22}\right),
\, e^{-\frac{6 i \pi}{11}}
\right),\;
\left(
\frac{2}{\sqrt{11}} \cos\!\left(\frac{5\pi}{22}\right),
\, e^{\frac{10 i \pi}{11}}
\right),\\[4pt]
&\left(
\frac{2}{\sqrt{11}} \sin\!\left(\frac{\pi}{11}\right),
\, 1
\right),\;
\left(
\frac{2}{\sqrt{11}} \sin\!\left(\frac{2\pi}{11}\right),
\, e^{-\frac{10 i \pi}{11}}
\right)
\,\bigg\} %\\
\times \bigg{\{} \left(\frac{1}2, 1\right)^{\otimes 2}, \left(\frac{1}2, i\right), \left(\frac{1}2, i^{-1}\right)\bigg{\}}\;.
\end{split}
\end{align}
The first factor comes from $\CD(11,-3)$, while the second comes from the decoupled ${\rm TFT}[\vec{k}]$. 

Using $\frac{2}{7} = \frac{1}{4- \frac{1}{2}}$, we find that
\begin{align}
\CD(7,2) \otimes {\rm TFT}[\vec{k}]\simeq  D(\vec{k})\textrm{ with }\vec{k}=(4,2)\;.
\end{align}
From \eqref{H and F of D(k)}, one can obtain the $A$-twisting limit of $\mathbf{HF}:=\{(\CH^{-1/2},\CF/\CF_{\alpha=0})\}$ for the $D((4,2))$ theory:
\begin{align}
\begin{split}
\mathbf{HF}^{A} &\textrm{ of } D((4,2))
\\
=    \bigg{\{}
 &\left( \frac{2}{\sqrt{7}} \cos\!\left(\frac{\pi}{14}\right), e^{-\frac{6 i \pi}{7}} \right), 
 \left( \frac{2}{\sqrt{7}} \cos\!\left(\frac{3 \pi}{14}\right), e^{-\frac{4 i \pi}{7}} \right), 
 \left( \frac{2}{\sqrt{7}} \sin\!\left(\frac{\pi}{7}\right), 1 \right)
\bigg{\}} \\
\times \bigg{\{} &\left(\frac{1}2, 1\right)^{\otimes 2}, \left(\frac{1}2, i\right), \left(\frac{1}2, i^{-1}\right)\bigg{\}}\;.
\end{split}
\end{align}
The first factor corresponds to $\CD(7,2)$, while the second arises from the decoupled ${\rm TFT}[\vec{k}]$. 

Taking the product of the first factor (from $\CD(11,-3)$), the second factor (from $\CD(7,2)$), and the last factor (from $U(1)_2$) yields the following set for the $\CT_{(7,11)}$ theory:
\begin{align}
\begin{split}
\mathbf{HF}^{A} &\textrm{ of } \CT_{(7,11)}
\\
=  \bigg\{\, &\left(
\frac{2}{\sqrt{11}} \cos\!\left(\frac{\pi}{22}\right),
\, e^{\frac{8 i \pi}{11}}
\right),\;
\left(
\frac{2}{\sqrt{11}} \cos\!\left(\frac{3\pi}{22}\right),
\, e^{-\frac{6 i \pi}{11}}
\right),\;
\left(
\frac{2}{\sqrt{11}} \cos\!\left(\frac{5\pi}{22}\right),
\, e^{\frac{10 i \pi}{11}}
\right),\\[4pt]
&\left(
\frac{2}{\sqrt{11}} \sin\!\left(\frac{\pi}{11}\right),
\, 1
\right),\;
\left(
\frac{2}{\sqrt{11}} \sin\!\left(\frac{2\pi}{11}\right),
\, e^{-\frac{10 i \pi}{11}}
\right)
\,\bigg\} \\
\times 
\bigg{\{}
 &\left( \frac{2}{\sqrt{7}} \cos\!\left(\frac{\pi}{14}\right), e^{-\frac{6 i \pi}{7}} \right), 
 \left( \frac{2}{\sqrt{7}} \cos\!\left(\frac{3 \pi}{14}\right), e^{-\frac{4 i \pi}{7}} \right), 
 \left( \frac{2}{\sqrt{7}} \sin\!\left(\frac{\pi}{7}\right), 1 \right)
\bigg{\}}
\\
\times \bigg{\{} &\left(\frac{1}{\sqrt{2}}, 1\right), \left(\frac{1}{\sqrt{2}}, i\right) 
\bigg{\}}\;.
\end{split}
\end{align}
This coincides with the set $(|S_{0\alpha}|, e^{2\pi i h_\alpha})$ of $M(7,11)$ as in Table \ref{table : dictionaries}.

\subsection{Comparison with the $\CT_{(2,2t+3)}$  by Gang-Kim-Stubbs }  For the case when $(P,Q)=(2,2t+3)$, the 3D $\CT_{(P,Q)}$ is (we choose $R=1,S=t+2$)
\begin{align}
\begin{split}
&\CT_{(2,2t+3)} \simeq  T_{\rm irred}[S^2 ((2,1),(2t+3, t+2),(3,1))]\;,
\\
&\simeq \CD(2t+3,t+2)  \otimes \frac{U(1)_2 \otimes U(1)_{-2}}{\mathbb{Z}_2^{\rm diag}}\simeq \CD(2t+3,t+2)\;.
\end{split}
\end{align}
Here, we use  the fact that  both $\CD(2,1)$ and $(U(1)_2 \otimes U(1)_{-2})/\mathbb{Z}_2^{\rm diag}$ are  trivial theories. Using $\frac{t+2}{2t+3} = \frac{1}{2 - \frac{1}{t+2}}$, the $\CD(2t+3,t+2)$ is given as
 \begin{align}
 D(\vec{k}) \simeq  \CD(2t+3,t+2)\otimes \textrm{TFT}[\vec{k}]  \textrm{ with }\vec{k}= (2,t+2)\;.
 \end{align}
 The decoupled topological field theory is, see \eqref{TFT-1} and \eqref{TFT-2}:
\begin{align}
{\rm TFT}[\vec{k}= (2,t+2)] \simeq  U(1)^2_{\CK} \textrm{ with mixed CS level }\CK= \begin{cases}
\begin{pmatrix}
0 & 2 \\ 2 &0
\end{pmatrix}, & t\in 2\mathbb{Z}
\\
\begin{pmatrix}
2 & 2 \\ 2 &0
\end{pmatrix}, & t\in 2\mathbb{Z}+1
\end{cases}
\end{align}
Recently, an abelian $\CN=2$ gauge theory description, denoted $\CT^{GKS}_{(2,2t+3)}$, of $\CT_{(2,2t+3)}$ was proposed in \cite{Gang:2023rei}:
\begin{align}
&\CT^{GKS}_{(2,2t+3)} = \left( \frac{(\CT_\Delta)^{\otimes t}}{U(1)^t_{K}}  +\textrm{monopole superpotentials}\right)\;,
\\
&\textrm{with mixed CS level : }
		K= 2	\begin{pmatrix}
			1 & 1 & 1 & \cdots & 1 & 1 \\
			1 & 2 & 2 & \cdots & 2 & 2 \\
			1 & 2 & 3 & \cdots & 3 & 3 \\
			\vdots & \vdots & \vdots & \ddots & \vdots & \vdots \\
			1 & 2 & 3 & \cdots & t-1 & t-1\\
			1 & 2 & 3 & \cdots & t-1 & t \\
		\end{pmatrix}\; ,
\end{align}
Here $\CT_\Delta$ is a free theory of a chiral multiplet with background CS level $-1/2$ for the $U(1)$ flavor symmetry \cite{Dimofte:2011ju}. 

We now claim that the two descriptions for $\CT_{(2,2t+3)}$ are actually equivalent.\footnote{In \cite{Comi:2023lfm}, they also found a dual description of the $\CT^{GKS}_{(2,2t+3)}$ theory, which is $T_{\rm full}[(t+1,1),(1,1),(1,1)]$.} One can check the following duality:
\begin{align}
\begin{split}
&\CT^{GKS}_{(2,2t+3)} \otimes \textrm{TFT}[(2,t+2)] \simeq D(\vec{k}= (2, t+2)) = \frac{T[SU(2)]}{(SU(2)_L)_{2} \times (SU(2)_R)_{t+2}}\;,
\end{split}
\end{align}
from various BPS partition function computations. For example, the superconformal index for $D(\vec{k})$ can be computed using \eqref{SCI for D(k)} and we find that
\begin{align}
\begin{split}
&\CI^{\rm sci}_{D(\vec{k}=(-2,-3))} (\eta, \nu=0)= \CI^{\rm sci}_{\CT^{GKS}_{(2,5)}}  (q,\eta, \nu=0)
\\
& =  1-q+\left(-\eta -\frac{1}{\eta }\right) q^{3/2}-2 q^2+\left(-\eta -\frac{1}{\eta }\right) q^{5/2}-2 q^3+\left(-\eta -\frac{1}{\eta }\right) q^{7/2}-2 q^4+\ldots
\\
&\CI^{\rm sci}_{D(\vec{k}=(-2,-4))}  (\eta, \nu=0) = \CI^{\rm sci}_{\CT^{GKS}_{(2,7)}} (q,\eta, \nu=0)
\\
& =  1-q+\left(-\eta -\frac{1}{\eta }\right) q^{3/2}-2 q^2-\eta  q^{5/2}+\left(\frac{1}{\eta ^2}-1\right) q^3+\left(\frac{1}{\eta }-\eta \right) q^{7/2}+\frac{q^4}{\eta ^2}+\ldots\;.
\end{split}
\end{align}
For the round 3-sphere partition function case, we have ($\simeq$ means equality up to an overall phase factor as defined in \eqref{phase factor ambiguity of S3b}):
\begin{align}
\begin{split}
&\left( \CZ^{\rm con}_{b=1} \textrm{ of } D(\vec{k} = (2,t+2)) \right) \simeq  \frac{1}{\sqrt{(2t+3)}} \sin \left (\frac{\pi}{2t+3} \right)\;,
\\
&\left( \CZ^{\rm con}_{b=1} \textrm{ of } \textrm{TFT}[\vec{k} = (2,t+2)]\right) \simeq   \frac{1}{\sqrt{|\det \CK|}} =\frac{1}2\;,
\\
&\left( \CZ^{\rm con}_{b=1} \textrm{ of } \CT^{GKS}_{(2,2t+3)}) \right) \simeq  \frac{2}{\sqrt{(2t+3)}} \sin \left (\frac{\pi}{2t+3} \right),
\end{split}
\end{align}
which again supports the proposed duality.

\section{Discussion and Future Directions}
In this paper, we provide an explicit field theory description of the 3D theory $\mathcal{T}_{(P,Q)}$ dual to the Virasoro minimal model $M(P,Q)$. Interestingly, the bulk theory exhibits very distinct IR phases—either gapped or $\mathcal{N}=4$ rank-0 SCFT—depending on whether the RCFT is unitary or not. The main results are presented in \eqref{T(PQ) from SFS} and \eqref{field theory for T(P,Q)}. 

\paragraph{Boundary Condition} In this paper, we compute various partition functions on closed 3-manifolds to test the bulk-boundary correspondence. To directly observe the boundary rational chiral algebra, one needs to consider the theory on an open manifold with an appropriate boundary condition \cite{Gadde:2013wq,Gadde:2013sca,Yoshida:2014ssa,Dimofte:2017tpi,Dedushenko:2017tdw}. Identifying the proper boundary condition that supports the Virasoro minimal models would be an interesting direction for future research.

\paragraph{Mirror RCFTs of Minimal Models} In 3D rank-0 SCFTs, there are two choices of topological twistings: $A$ and $B$ twisting. Our theory $\mathcal{T}_{(P,Q)}$ is expected to support the Virasoro minimal model at the boundary under one of these topological twistings. The other choice of twisting generally supports a different rational chiral algebra at the boundary. Understanding the mirror dual rational chiral algebras of the Virasoro minimal models would be a valuable avenue for further study.

\paragraph{Other Minimal Models} Recently, 3D dual theories for some supersymmetric $\mathcal{N}=1$ minimal models have been proposed \cite{Baek:2024tuo}. Extending our work to other classes of minimal models, including these supersymmetric cases, would be an intriguing direction for future research. Some progress in this direction will be reported in \cite{BGK:2025}.

\acknowledgments{We would like to thank  Yuji Tachikawa for the useful discussion.
	The work of DG and HK is supported in part by the National Research Foundation of Korea grant  NRF-2022R1C1C1011979. DG also acknowledges support by Creative-Pioneering Researchers Program through Seoul National University. }

%%%%%%%%%%%%%%%%%%%%%%%%%%%%%%%%%%%%%%%%%%%%%%%%%%%%%%%%%%%%%%%%%%%%%%%%
\appendix
%%%%%%%%%%%%%%%%%%%%%%%%%%%%%%%%%%%%%%%%%%%%%%%%%%%%%%%%%%%%%%%%%%%%%%%%
\section{BPS partition functions of $D(\vec{k})$ } \label{appendix : Z[D(p,q)]}

The $T[SU(2)]$ theory, which is a basic building block of the $D(\vec{k})$ theory,  is a 3D $\CN=4$ SQED with $N_f=2$ (see Table \ref{T[SU(2)]}). 
\begin{table}[ht]
	\begin{center}
		\begin{tabular}{|c|c|c|c|c|}
			\hline
			Chiral multiplet  &  $U(1)_{\rm gauge }$ & $R_{\nu=0}$ &  $A$ & $F$
			\\
			\hline \hline
			$(\Phi_1,\Phi_2) $ & $(+1, -1)$ & $\frac{1}2$ &  $\frac{1}2$ & $(+1, -1)$
			\\
			\hline
			$(\Phi_3, \Phi_4)$ & $(+1, -1)$  & $\frac{1}2$ & $\frac{1}2$ & $(-1,+1)$
			\\
			\hline
			$\Phi_0$ & $0$  & $1$ & $-1$ & $0$
			\\
			\hline
		\end{tabular}
	\end{center}
	\caption{Matter contents of the $T[SU(2)]$ theory  in terms of $\CN=2$ chiral multiplets. $(\Phi_1, \Phi_2)$ and $(\Phi_3, \Phi_2)$ form two $\CN=4$ hypermultiplets transforming as $\mathbf{2}$ under the $SU(2)_L$ flavor symmetry, with their Cartan denoted by $F$, normalized as $F\in \mathbb{Z}$. The theory has a $U(1)$ topological symmetry associated with the $U(1)$ gauge field, which is enhanced to $SU(2)_R$ in the IR. $\CN=4$ vector multiplet contains a neutral chiral multiplet $\Phi_0$. The theory has a superpotential $\CW \propto   \Phi_1 \Phi_0 \Phi_2 - \Phi_3 \Phi_0 \Phi_4  $. }
	\label{T[SU(2)]}
\end{table}
In terms of an $\CN=2$ subalgebra, the theory possesses $SU(2)_L\times SU(2)_R\times U(1)_A$ flavor symmetry that commutes with the subalgebra.  Various SUSY partition functions of the $T[SU(2)]$ theory and its variants have been computed in various literature. To ensure self-containment, we reproduce these computations here.

\paragraph{Squashed 3-sphere partition function \cite{Kapustin:2009kz,Jafferis:2010un,Hama:2010av}} The squashed 3-sphere partition function $\mathcal{Z}^{S^3_b}$ of the $S$-duality wall theory is 
\begin{align}
\begin{split}
&\CZ^{S^3_b}_{T[SU(2)]} (M_1, M_2; M , \nu) = \int \frac{d Z}{\sqrt{2\pi \hbar}}  \CI^{\hbar}_{T[SU(2)]} (Z, M_1, M_2; W),\; \textrm{  where }
\\
&\CI^{\hbar}_{T[SU(2)]}:= \exp \left(\frac{Z^2+M_1^2 +2M_2 Z}{\hbar}\right) \psi_\hbar \left( -W +(i \pi +\frac{\hbar}2 )\right) 
\\
&\qquad \qquad \quad \times   \prod_{\epsilon_1, \epsilon_2  \in \{\pm 1 \} } \psi_\hbar \left( \epsilon_1  Z + \epsilon_2 M_1+\frac{W}2 +(\frac{i \pi}2  +\frac{\hbar}4 )\right)\bigg{|}_{W:=M +(i \pi +\hbar/2) \nu}\;,
\end{split}
\end{align}
We define $\hbar :=2\pi i b^2$, where $b$ is the squashing parameter of the squashed 3-sphere $S^3_b$ in \eqref{squashed 3-sphere}.

The partition function has following  phase factor ambiguity
\begin{align}
\exp \left(i \pi (b^2+b^{-2}) \mathbb{Q} + i \pi \mathbb{Q}\right) \;, \label{phase factor ambiguity of S3b}
\end{align}
which depends on the background CS levels for $R$-symmetry and flavor symmetries, decoupled invertible TQFT, 3-manifold framing and so forth.  
The partition function depends on following parameters: 
\begin{align}
\begin{split}
&M_1/ M_2 \;:\; \textrm{(rescaled) real masses for the  the Cartan $U(1)$s of $SU(2)_L/SU(2)_R$}\;,\\
&(M,\nu) \;:\; \textrm{((rescaled) real mass, R-symmetry mixing in \eqref{R-symmetry mixing}) for the $U(1)_A$}\;. 
\end{split}
\end{align}
The rescaled real mass $M$ is $ b\times (\textrm{real mass})$ and the squashed 3-sphere partition has $b\leftrightarrow b^{-1}$ symmetry  when the (unrescaled) real masses and $\nu$ are fixed.  The special function $
\psi_\hbar$ in the integrand is called quantum dilogarithm (Q.D.L) function, which is defined by \cite{Faddeev:1993rs} 
\begin{align}
\psi_\hbar (Z):= \begin{cases} 
\prod_{r=1}^{\infty} \frac{1-q^r e^{-Z}}{1-\widetilde{q}^{-r+1} e^{-\widetilde{Z}}}\;, \quad \textrm{if $|q|<1$}
\\
\prod_{r=1}^{\infty} \frac{1-\widetilde{q}^r e^{-\widetilde{Z}}}{1-q^{-r+1} e^{-Z}}\;, \quad \textrm{if $|q|>1$}
\end{cases}
\end{align}
with
\begin{align} 
q=e^{2\pi i b^2}, \quad \widetilde{q}:=e^{2\pi i b^{-2}}, \quad \widetilde{Z}= Z/b^2\;.
\end{align}
The $\psi_\hbar (Z)$ computes the squashed 3-sphere partition function of the $\CT_\Delta$ theory \cite{Dimofte:2011ju}, a massless free theory of a single  $\CN=2$ chiral multiplet with background CS level $-1/2$ for the $U(1)$ flavor symmetry. The $Z$ is $M+(i \pi+\frac{\hbar}2 R(\Phi))$ with the  rescaled real mass $M$ for the $U(1)$ flavor symmetry and the R-charge  $R(\Phi)$ of  the chiral field.  
The self-mirror property of the $T[SU(2)]$ theory implies that
\begin{align}
\CZ^{S^3_b}_{T[SU(2)]} (M_1, M_2 ; M, \nu) \simeq \CZ^{S^3_b}_{T[SU(2)]} (M_2, M_1;-M, -\nu)\;.
\end{align}
Here $\simeq$ means the equality modulo a phase factor of the form in \eqref{phase factor ambiguity of S3b}. 
Using the $T[SU(2)]$ partition function, the squashed 3-sphere partition function of $D(\vec{k} )$ in Figure \ref{fig:quiver} can be computed as follows
\begin{align}
\begin{split}
&\CZ^{S^3_b}_{D(\vec{k})} (M, \nu) 
\\
&= \int \left( \prod_{I=1}^\sharp\frac{ \Delta(M_I) dM_I}{\sqrt{2\pi \hbar}}   \exp \left(\frac{k^{(I)}M_I^2}{\hbar}\right) \right)    \left( \prod_{I=1}^{\sharp-1}\CZ^{S^3_b}_{T[SU(2)]} (M_I, M_{I+1};M, \nu) \right)\;.
\end{split} \label{S3b for D(k)}
\end{align}
The $\Delta (M)$ is the contribution from a $SU(2)$ vector multiplet,
\begin{align}
\Delta (M) = 2\sinh (M) \sinh (2\pi i M /\hbar)\;.
\end{align}
The partition function for $D(\vec{k})$ drastically  simplifies when $b=1$ and $\nu=0$. For $T[SU(2)]$ theory, the partition function becomes  \cite{Nishioka:2011dq,Gang:2021hrd}
\begin{align}
\CZ^{S^3_{b=1}}_{T[SU(2)]} (M_1, M_2; M=0, \nu=0) \simeq \frac{1}2  \frac{\sin \left( \frac{M_1 M_2}{\pi}\right)}{\sinh(M_1)\sinh(M_2)}\;.
\end{align}
For $D(\vec{k})$ theory, 
\begin{align}
\begin{split}
&\CZ^{S^3_{b=1}}_{D(\vec{k})} (M=0, \nu=0) 
\\
&\simeq 2 \int \left( \prod_{I=1}^{\sharp} \frac{dM_I}{2\pi }  \exp \left(\frac{k^{(I)}M_I^2}{2\pi i }\right)\right) \sinh(M_1)\sinh(M_{\sharp}) \prod_{I=1}^{\sharp-1} \sin\left(\frac{M_I M_{I+1}}\pi \right)\;,
\\
&\simeq  \frac{1}{\sqrt{2^{\sharp-2}|p|}} \sin\left( \frac{\pi}{|p|}\right)\;. \label{S3 for D(k)}
\end{split}
\end{align}
The integral in the middle line is simply a sum of Gaussian integrals and can be easily evaluated to obtain the final answer. Interestingly, the final result can be expressed as a very simple function of $p$, which is related to $\vec{k}$ as shown in \eqref{p/q and k}. In the computation, we use the following identity:
\begin{align}
\begin{split}
&\det \overline{\CK} (\vec{k}) = |p|\;, 
\\
&\textrm{ where } \overline{\CK}_{IJ} := \begin{cases}
1, & |I-J| =1
\\
k^{(I)}, & I=J
\\
0, & \textrm{otherwise}
\end{cases}\;\quad \quad  ( I,J=1,\ldots, \sharp)
\end{split} \label{det of Kbar}
\end{align}
\paragraph{Superconformal index \cite{Kim:2009wb,Imamura:2011su,Dimofte:2011py}} The generalized superconformal index $\CI^{\rm sci}_{T[SU(2)]}$ for the $T[SU(2)]$ theory is $(\tilde{\eta}^2 := \eta)$
\begin{align}
\begin{split}
& \CI^{\rm sci}_{T[SU(2)]} (m_1, u_1, m_2, u_2; \eta, \nu) 
\\
&=  \sum_{m}\oint_{|u|=1}\frac{d u}{2\pi i u}  u^{2(m+m_2)} u_2^{2m} u_1^{2m_1}  \CI_\Delta (0, -q^{1/2}\tilde{\eta}^{-2})   
\\
& \quad  \times \prod_{\epsilon_1, \epsilon_2 \in \{\pm 1 \}} \CI_\Delta (\epsilon_1 m+ \epsilon_2 m_1,  \tilde{\eta} u^{\epsilon_1} u^{\epsilon_2}_1 (-q^{1/2})^{1/2})  \big{|}_{\tilde{\eta} \rightarrow (- \eta q^{1/2})^{\nu/2}} \;,
\\
&=  \sum_{m}\oint_{|u|=1}\frac{d u}{2\pi i u}  (-q^{1/2})^{m+m_2} u^{2(m+m_2)} u_1^{2m_1} u_2^{2m} \CI_\Delta (0, -q^{1/2}\tilde{\eta}^{-2})     \CI_\Delta (m+m_1, -q^{1/2} \tilde{\eta} u u_1)
\\
&  \quad \quad  \CI_\Delta (m-m_1, -q^{1/2} \tilde{\eta} u u_1^{-1})  \CI_\Delta (-m+m_1,  \tilde{\eta} u^{-1} u_1)  \CI_\Delta (-m-m_1, \tilde{\eta} u^{-1}u_1^{-1})\big{|}_{\tilde{\eta} \rightarrow (- \eta q^{1/2})^{\nu/2}} \;.
\end{split}
\end{align}
In the middle, we changed the integral variable $u$ to $u (-q^{1/2})^{1/2}$, which corresponds to adjusting the mixing between the $U(1)$ R-symmetry and the gauge $U(1)$ symmetry, i.e.,
\begin{align}
R_{\nu} \rightarrow \tilde{R}_\nu := R_\nu + \frac{1}2 G\;,
\end{align}
where $G$ is the gauge charge. Since the index counts gauge-invariant operators, it remains unaffected by the mixing. In practice, the last expression is much easier to handle using Mathematica.
The index depends on the following parameters: 
\begin{align}
\begin{split}
&(m_1, u_1)/(m_2, u_2)\;:\;\textrm{(monopole flux, fugacity) for the  the Cartans of $ SU(2)_L/SU(2)_R$}\;,\\
&(\eta, \nu)  \;:\; \textrm{fugacity and R-symmetry mixing parameter for  the $U(1)_A$ symmetry}\;. 
\end{split}
\end{align}
Here the tetrahedron index $\CI_\Delta (m,u)$ is defined as \cite{Dimofte:2011py}
\begin{align}
\begin{split}
&\CI_\Delta (m,u) := \prod_{r=0}^\infty \frac{1-q^{r-\frac{1}2 m+1}u^{-1}}{1-q^{r- \frac{1}2 m} u} =\sum_{e \in \mathbb{Z}}\CI_\Delta^{c} (m, e) u^e,  \;\; 
\\
&\textrm{where  } \CI_\Delta^c (m,e) = \sum_{n=\lfloor e \rfloor }^\infty \frac{(-1)^n q^{\frac{1}2 n (n+1)-(n+\frac{1}2 e)m}}{(q)_n (q)_{n+e}}\;.
\end{split}
\end{align}
It computes the generalized superconformal index of the $\CT_\Delta$ theory with the R-charge choice $R(\Phi)=0$ where $(m,u)$ are (background monopole flux, fugacity) for the $U(1)$ flavor symmetry. At  general $R$-charge choice, the index becomes $\CI_\Delta(m, u(-q^{1/2})^{R(\Phi)})$.  As a consistency check for the formula, one can confirm the following self-mirror  property in $q$-expansion
\begin{align}
\CI^{\rm sci}_{T[SU(2)]} (m_1, u_1, m_2, u_2 ;\eta, \nu )   = \CI^{\rm sci}_{T[SU(2)]} (m_2, u_2, m_1, u_1 ;\eta^{-1}, -\nu)\;.
\end{align}
 Using the $T[SU(2)]$ index, the superconformal index of the  $D(\vec{k} )$ theory can be computed as follows:
\begin{align}
\begin{split}
&\CI^{\rm sci}_{D(\vec{k})} (\eta , \nu) 
\\
&= \sum_{m_1, \ldots, m_m \in \mathbb{Z}_{\geq 0}} \oint  \prod_{I=1}^\sharp \left( \frac{du_I}{2\pi i u_I}  \Delta (m_I, u_I)  u_I^{2k^{(I)}m_I} \right)   \left( \prod_{I=1}^{\sharp-1}\CI^{\rm sci}_{T[SU(2)]} (m_I, u_I, m_{I+1}, u_{I+1};\eta, \nu)\right)\;. \label{SCI for D(k)}
\end{split}
\end{align}
Here $\Delta (m,u)$ is the contribution from an $SU(2)$ vector multiplet
\begin{align}
\begin{split}
&\Delta (m,u) = \frac{1}{\textrm{Sym}(m)} (q^{m/2}u -q^{-m/2}u^{-1})  (q^{m/2}u^{-1} -q^{-m/2}u) 
\\
&\textrm{with }\textrm{Sym}(m) := \begin{cases} 2, \; m=0 \\ 1 , \; m\neq 0 \end{cases}\;.
\end{split}
\end{align}
One can also check the following
\begin{align}
\begin{split}
&\CI^{\rm sci}_{T[SU(2)]/(SU(2)_R)_{\pm 1}} (m_1, u_1; \eta, \nu) 
\\
&= \sum_{m_2 \in \mathbb{Z}_{\geq 0} } \oint \frac{du_2}{2\pi i u_2} \Delta (m_2, u_2) u_2^{\pm 2 m_2} \CI^{\rm sci}_{T[SU(2)]} (m_1, u_1, m_2, u_2 ; \eta, \nu)= u_1^{\mp 2 m_1}\;. \label{SCI for TSU2/SU(2)}
\end{split}
\end{align}
It provides a non-trivial check for the IR duality corresponding to the first move in Figure \ref{fig:Kirby-moves}. 
\paragraph{Twisted partition functions \cite{Benini:2015noa,Benini:2016hjo,Closset:2016arn,Closset:2018ghr}} Now let us compute the twisted partition functions $\CZ^{\CM_{g,p}}$ \eqref{twisted ptns} of the $D(\vec{k})$ theory. For the computation, we first consider the integrand of the squashed 3-sphere partition function \eqref{S3b for D(k)} in an  asymptotic limit $\hbar \rightarrow 0$ \cite{Gang:2019jut}:
\begin{align}
\begin{split}
&\log \CI^\hbar_{D(\vec{k})} \xrightarrow{\quad \hbar \rightarrow 0 \quad }  \frac{1}\hbar \CW^{(\vec{k})}_0 + \CW^{(\vec{k})}_1 +\ldots\;,
\\
&\CW_0^{(\vec{k})}(\vec{Z}, \vec{M}; M, \nu) = \left( \sum_{I=1}^{\sharp} (\pm 2\pi i M_I +k^{(I)} M_I^2)  \right)+\sum_{I=1}^{\sharp-1} \CW_0^{T[SU(2)]} (Z_I, M_I, M_{I+1};M, \nu)\;,
\\
&\textrm{where } \CW_0^{T[SU(2)]} (Z,M_1, M_2; M, \nu)= Z^2+M_1^2+2 M_2 Z  + \textrm{Li}_2 (e^{M+i \pi \nu}) 
\\
& \qquad \qquad \qquad \qquad \qquad \qquad \qquad \qquad + \sum_{\epsilon_1 , \epsilon_2 \in \{\pm1 \}}      \textrm{Li}_2 (-e^{- \epsilon_1 Z-\epsilon_2 M_1 -\frac{(M+i \pi (\nu-1) )}2})  \;,
 \\
 &\CW^{(\vec{k})}_1 (\vec{Z},\vec{M};M, \nu)  =\sum_{I=1}^{\sharp} \log (\sinh(M_I))+\sum_{I=1}^{\sharp-1} \CW_1^{T[SU(2)]}  (Z_I, M_I, M_{I+1};M, \nu)  \textrm{ with } 
 \\
 &\CW_1^{T[SU(2)]} (Z, M_1, M_2) = - \frac{\nu}2 \log (1+e^{M+i \pi \nu}) 
 \\
 & \qquad \qquad \qquad \qquad  \qquad \;\; + \frac{\nu-1}{4}\sum_{\epsilon_1 , \epsilon_2 \in \{\pm1 \}}      \log(1+e^{- \epsilon_1 Z-\epsilon_2 M_1 -\frac{(M+i \pi (\nu-1) )}2})  \;.
\end{split}
\end{align}
We use the following asymptotic behavior of Q.D.L.,
\begin{align}
\log \psi_\hbar (Z) \xrightarrow{\quad \hbar \rightarrow 0 \quad } \frac{\textrm{Li}_2(e^{-Z})}\hbar - \frac{1}2 \log (1-e^{-Z})+ \ldots\;.
\end{align}
Then, the Bethe-vacua of the $D(\vec{k})$ theory are obtained as follows $(I=1,\ldots, (\sharp -1)$ while $J= 1, \ldots, \sharp)$
\begin{align}
\begin{split}
&\CS^{\rm BE}_{D(\vec{k})} (M, \nu)= \big{\{}\vec{z}, \vec{m} \;:\; \exp (\partial_{Z_I} \CW_0^{(\vec{k})})\big{|}_{*} =  \exp (\partial_{M_J} \CW_0^{(\vec{k})})\big{|}_{*}  = 1, \; m_J^2 \neq 1  \big{\}}/\mathbf{W}\;,
\\
&\textrm{with } * : Z_I \rightarrow \log z_I, \; M_J \rightarrow \log m_J\;.
\end{split}
\end{align} 
Here $\mathbf{W}$ denotes the Weyl subgroup  $\mathbb{Z}_2^{\sharp}$  of the $SU(2)^{\sharp}$ gauge symmetry, which acts on the Bethe-vacua as
\begin{align}
\mathbf{W} \;:\; m_J \rightarrow 1/m_J \;\; \textrm{for each $J=1, \ldots, \sharp$}\;.
\end{align}
Handle gluing $\CH$ and fibering operator $\CF$ of the $D(\vec{k})$ theory are ($\vec{X}:= (\vec{Z},\vec{M})$)
\begin{align}
\begin{split}
&\CH_{D(\vec{k})} (\vec{z}, \vec{m};M , \nu) = \frac{e^{i \delta }} {|\mathbf{W}|^{2}}\left( \det_{A,B} \left(  \partial_{X_A} \partial_{X_B}  \CW^{(\vec{k})}_0 \right) \right) \exp\left( -2\CW_1^{(\vec{k})} \right) \bigg{|}_{*}\;,
\\
&\CF_{D(\vec{k})} (\vec{z},\vec{m}; M, \nu) = \exp\left(- \frac{\CW^{(\vec{k})}_0 - \vec{X}\cdot \partial_{\vec{X}} \CW^{(\vec{k})}_0 - M \partial_M \CW^{(\vec{k})}_0}{2\pi i } \right) \bigg{|}_{*}\;. \label{H and F of D(k)}
\end{split}
\end{align}
Here $|\mathbf{W}| = 2^{\sharp}$ and $e^{i \delta}$ is a phase factor which is sensitive to the subtle overall phase factor in \eqref{phase factor ambiguity of S3b}. We fix the phase ambiguity by requiring that
\begin{align}
\mathcal{Z}^{\CM_{g=0,p=0}}_{D(\vec{k})} (M=0, \nu=\pm 1) = \sum_{(\vec{z},\vec{m})\in \CS_{\rm BE}} \CH_{D(\vec{k})}^{-1}=1\;.
\end{align}
For $D(\vec{k})$ theory with $\vec{k}= (k_1, k_2)$, the twisted partition functions are studied in \cite{Gang:2021hrd}. There are $2 \times (|k_1 k_2 -1|-1)$ Bethe-vacua whose handle gluing operators 
\begin{align}
\{\CH_{D(\vec{k} = (k_1, k_2))}  (\vec{z},\vec{m})\;:\; {(\vec{z}, \vec{m}) \in \CS_{\rm BE}}(M=0, \nu = \pm 1) \} =  \bigg{\{}\frac{|k_1 k_2-1|}{\sin^2 \left( \frac{\pi n }{k_1 k_2 -1}\right)}^{\otimes 2} \bigg{\}}_{n=1}^{|k_1 k_2-1|-1}\;.
\end{align}
Further, one can check that the set  $\mathbf{HF}:=\{ (\CH^{-1/2},\CF ) \;:\; (\vec{z},\vec{m}) \in \CS_{\rm BE} \}  $ of $D(\vec{k}=(k_1, k_2))$ has following factorization properties
\begin{align}
\mathbf{HF} =  \widetilde{\mathbf{HF}} \times \begin{cases}\{ (\frac{1}2,1),(\frac{1}2,1),(\frac{1}2,1),(\frac{1}2,-1)\}, \quad & \textrm{$k_1$ and $k_2$ are both even}  \;,
\\
 \{ (\frac{1}2,1),(\frac{1}2,1),(\frac{1}2,i ),(\frac{1}2,i^{-1} )\}, & \textrm{one of  them is odd}
\\
\{ (\frac{1}{\sqrt{2}},1),(\frac{1}{\sqrt{2}}, i)\} \textrm{ or } \{ (\frac{1}{\sqrt{2}},1),(\frac{1}{\sqrt{2}}, i^{-1})\},& \textrm{both are odd}
\end{cases} \label{factorization of HF}
\end{align}
%
%Here $\mathbb{Z}_2^{[1]} \times \mathbb{Z}_2^{[1]}$ is the two $\mathbb{Z}_2$ 1-form symmetry of the $D(\vec{k})$ theory, which is originated from the center subgroup of the gauge symmetry, $SU(2)\times SU(2)$.  The 1-form symmetry acts on the Bethe-vacua in the following way
%
%\begin{align}
%\mathbb{Z}_2^{[1]} \times \mathbb{Z}^{[1]} \;:\; (m_1, m_2) \rightarrow (\pm m_1, \pm %m_2)
%\end{align}
%
%When  $k_1$ and $k_2$ both are odd, for each Bethe-vacuum there is a  $\mathbb{Z}^{[1]}_2$ subgroup of the 1-form symmetry which trivially acts on the Bethe-vacuum. 
We will understand the above factorization pattern by analyzing the decoupled TQFT ${\rm TFT}[\vec{k}]$ in the next section. 
\section{Decoupled TQFT $\textrm{TFT}(\vec{k})$} \label{appendix: decoupled TQFT}
From the 't Hooft anomaly \eqref{tHooft anomaly} of the $(\mathbb{Z}_2)^{\sharp}$ 1-form symmetry in $D(\vec{k})$, we expect that the $D(\vec{k})$ theory in the IR contains a decoupled topological field theory which has the same anomaly. The simplest choice is the $U(1)^{\sharp}_{\CK}$ theory with the mixed CS term $\CK$ of the following form:
\begin{align}
\frac{\CK}{2} = ( \overline{\CK} \textrm{ in \eqref{det of Kbar}}) \textrm{ (mod $2$)}\;. \label{decoupled TQFT-0}
\end{align}
This implies that
\begin{align}
\begin{split}
&\det (\CK/2) = \det (\overline{\CK}) = |p|  \; (\textrm{mod $2$})\;,
\\
&\Rightarrow  |\det (\CK)| = \begin{cases} \in 2^{\sharp } \times (2 \mathbb{Z}_{\geq 0}+1), & \textrm{odd $p$}
\\
\in 2^{\sharp } \times (2 \mathbb{Z}_{\geq 0}), & \textrm{even $p$} \label{decoupled TQFT-1}
\end{cases}
\end{split}
\end{align}
On the other hand, for the  $U(1)^{\sharp }_{\CK}$ theory  $(\textrm{when $\CK$ is non-degenerate})$
\begin{align}
\begin{split}
& \det (\CK) = (\textrm{the number of ground states on $\mathbb{T}^2$, i.e. Bethe-vacua}) 
\\
&\leq 2^{\sharp}\;. \label{decoupled TQFT-2}
 \end{split}
\end{align}
since all the Bethe-vacua of the decoupled TQFT are connected to each other by an action of the $(\mathbb{Z}_2)^{\sharp}$ 1-form symmetry. Combining \eqref{decoupled TQFT-1} and \eqref{decoupled TQFT-2}, the $\det \CK$ is determined as follows:
\begin{align}
\det (\CK) = \begin{cases} 2^{\sharp }, & \textrm{odd $p$}\\ 0 , & \textrm{even $p$} \end{cases} \label{decoupled TQFT-3}
\end{align}
\paragraph{ $\sharp=2$ and $k_1 , k_2\in 2\mathbb{Z}$} In the case,  $p$ is always odd and the possible candidates for $\CK$ are ($n\in \mathbb{Z}$)
\begin{align}
\CK = \begin{pmatrix} 4n &  \pm 2 \\ \pm 2 & 0  \end{pmatrix} \textrm{ or } \begin{pmatrix} 0 &  \pm 2 \\ \pm 2 & 4n  \end{pmatrix} 
\end{align}
But all the $\CK$s are actually equivalent to 
\begin{align}
\CK \simeq   \begin{pmatrix} 0 &    2 \\  2 & 0  \end{pmatrix} \;.
\end{align}
Here $\CK \simeq \CK'$ is an equivalence between $U(1)^{\sharp}_\CK$ and  $U(1)^{\sharp}_{\CK'}$ up to a following redefinition of gauge fields $\vec{A}$
\begin{align}
\vec{A} \rightarrow M \cdot \vec{A}\textrm{ with }M\in GL(2\sharp ,\mathbb{Z})\;.
\end{align}
We choose $GL(2\sharp ,\mathbb{Z})$ instead of  $GL(2\sharp ,\mathbb{R})$ to preserve the monopole charge quantization.  
More explicitly, the equivalence relation is 
\begin{align}
\CK \simeq \CK' \textrm{ if } \CK'= M^T \CK M \textrm{ with an $M\in GL(2\sharp ,\mathbb{Z})$}\;.
\end{align}
So, in the case, the decoupled TQFT is uniquely determined
\begin{align}
\textrm{TFT}[(k_1,k_2 )] \simeq  U(1)^2_{\CK}  \textrm{ with } \CK= \begin{pmatrix} 0 &    2 \\  2 & 0  \end{pmatrix}\;,\label{TFT-1}
\end{align}
which is the toric-code TQFT. It  has the  set of $\mathbf{HF}:= \{ |S_{0\alpha}|, \exp(2\pi i h_\alpha) \}$ as follows \footnote{For the $U(1)^{\sharp}_\CK$ theory with non-degenerate $\CK$, there are $\det \CK$ simple objects (or Bethe-vacua), $\alpha=0,\ldots , \det \CK-1$, which  are solutions of the Bethe equations, $\CS_{\rm BE}:=\{\vec{z}\;: \prod_{I=1}^{\sharp}z_I^{\CK_{IJ}}=1, \;  \textrm{ for }J=1,\ldots, \sharp\}$. The $S_{0\alpha}$ of  Bethe-vacuum
	 $\vec{z}_\alpha \in \CS_{\rm BE}$ is $1/\sqrt{|\det \CK|}$ for all $\alpha $ and the $e^{2\pi i h_\alpha}$ is given by the fibering operator $\CF=\exp(\frac{\sum_{I,J}\CK_{IJ}Z_I Z_J}{4\pi i })$ with $\vec{Z} = \log \vec{z}_\alpha$. }
\begin{align}
(\mathbf{HF}  \textrm{ of }\textrm{TFT}[(k_1, k_2)])=  \bigg{\{}(\frac{1}2 ,1)^{\otimes 3} , (\frac{1}2,-1) \bigg{\}}\;,
\end{align}
which explains the corresponding factorization property in \eqref{factorization of HF}.

\paragraph{ $\sharp=2$ and $k_1 \neq k_2 \;(\textrm{mod }2)$ } In the case, $p$ is always odd and there is only one consistent choice of $\CK$ up to the equivalence, which is 
\begin{align}
\textrm{TFT}[(k_1,k_2)] \simeq  U(1)^2_{\CK}  \textrm{ with } \CK= \begin{pmatrix} 2 &    2 \\  2 & 0  \end{pmatrix}\;, \label{TFT-2}
\end{align}
whose $\mathbf{HF}:= \{ |S_{0\alpha}|, \exp(2\pi i h_\alpha) \}$ is 
\begin{align}
(\mathbf{HF}  \textrm{ of }\textrm{TFT}[(k_1, k_2)])=  \bigg{\{}(\frac{1}2 ,1)^{\otimes 2} , (\frac{1}2,i), (\frac{1}2,\frac{1}{i}) \bigg{\}}\;,
\end{align}
which explains the corresponding factorization property in \eqref{factorization of HF}.

\paragraph{ $\sharp=2$ and $k_1 , k_2 \in 2\mathbb{Z}+1$ } In the case, $p$ is always even and the possible candidates for the $\CK$ are
\begin{align}
\CK \simeq 2 \begin{pmatrix} a & a \\ a & a\end{pmatrix}  \simeq  2 \begin{pmatrix} 0 & 0 \\ 0 & a\end{pmatrix}  \textrm{ with $a\in 2\mathbb{Z}+1$}\;.
\end{align}
Ignoring the  1st gauge field $A_1$ which does not appear in the action, the decoupled TQFT  is nothing but the $U(1)_{2a}$ theory. From the constraint in \eqref{decoupled TQFT-2}, which is $|2a|\leq 2^2$ for our case, the possible values of $a$ are $\pm 1$. Thus, the decoupled TQFT is $U(1)_{\pm 2}$ theory  
whose $\mathbf{HF}:= \{ |S_{0\alpha}|, \exp(2\pi i h_\alpha) \}$ is 
\begin{align}
(\mathbf{HF}  \textrm{ of }\textrm{TFT}[(k_1, k_2)])=  \bigg{\{}(\frac{1}{\sqrt{2}} ,1), (\frac{1}{\sqrt{2}}, i^{\pm 1} ) \bigg{\}}\;,
\end{align}
which explains the corresponding factorization property in \eqref{factorization of HF}.

For higher values of $\sharp \geq 3$, the analysis becomes more complicated, and we could not uniquely determine $\CK$. From the factorization properties of the set $\mathbf{HF}$ for several $D(\vec{k})$s, we observe that
\begin{align}
(\textrm{the number of Bethe-vacua of the decoupled TQFT}) = \begin{cases} 2^{\sharp}, & \textrm{odd }p
\\
2^{\sharp-1}, & \textrm{even }p \label{GSD for decouple TQFT}
\end{cases}
\end{align}
For odd $p$, this is compatible with \eqref{decoupled TQFT-2}  and \eqref{decoupled TQFT-3}.  For  abelian CS theory, it is  generally true that
 \begin{align}
  (\textrm{3-sphere partition function}) = 1/\sqrt{(\textrm{the number of Bethe-vacua})}\;.
 \end{align}
 Thus, from \eqref{GSD for decouple TQFT}, we have
 \begin{align}
 \begin{split}
 &|(\mathcal{Z}^{\rm con} \textrm{ of } \textrm{TFT}(\vec{k} )) | = \begin{cases} 2^{-\sharp/2}, & p \in 2\mathbb{Z}+1
 \\
 2^{(1-\sharp)/2}, & p \in 2\mathbb{Z} 
 \end{cases}
 \end{split}
 \end{align}
 Combined  with \eqref{S3 for D(k)}, we obtain
 \begin{align}
 \begin{split}
 &|(\mathcal{Z}^{\rm con} \textrm{ of } \CD(p,q)) | = \frac{|(\mathcal{Z}^{\rm con} \textrm{ of } D(\vec{k})) |}{|(\mathcal{Z}^{\rm con} \textrm{ of } \textrm{TFT}(\vec{k} ))  |}  = \begin{cases} \frac{2}{\sqrt{|p|}} \sin\left( \frac{\pi}{|p|}\right) , & p \in 2\mathbb{Z}+1
 \\
 \sqrt{\frac{2}{|p|}} \sin\left( \frac{\pi}{|p|}\right), & p \in 2\mathbb{Z} 
 \end{cases}
 \end{split} \label{S3 D(p,q)}
 \end{align}
 Note that the partition function is independent on the choice of $\vec{k}$ as expected from $i)$ in \eqref{Two properties of D(p,q)}. 
 
 For even $p$, a  $\mathbb{Z}_2$ subgroup of the UV  $(\mathbb{Z}_2)^{\sharp}$ 1-form symmetry is absent in the decoupled $\textrm{TFT}(\vec{k})$, which captures the anomaly of the 1-form symmetry. This implies that 
 \begin{align}
 \textrm{For }p\in 2\mathbb{Z}, \;\; \CD(p,q) \textrm{ has a non-anomalous $\mathbb{Z}_2$ 1-form symmetry}, \label{Z2 in even p}
 \end{align}
which can be identified with the  absent $\mathbb{Z}_2$ subgroup.
\bibliographystyle{ytphys}
\bibliography{ref}

@inproceedings{Dedushenko:2017tdw,
    author = "Dedushenko, Mykola and Gukov, Sergei and Putrov, Pavel",
    title = "{Vertex algebras and 4-manifold invariants}",
    booktitle = "{Nigel Hitchin's 70th Birthday Conference}",
    eprint = "1705.01645",
    archivePrefix = "arXiv",
    primaryClass = "hep-th",
    reportNumber = "CALT-TH-2017-008",
    doi = "10.1093/oso/9780198802013.003.0011",
    volume = "1",
    pages = "249--318",
    month = "5",
    year = "2017"
}

@article{Hsin:2018vcg,
	author = "Hsin, Po-Shen and Lam, Ho Tat and Seiberg, Nathan",
	title = "{Comments on One-Form Global Symmetries and Their Gauging in 3d and 4d}",
	eprint = "1812.04716",
	archivePrefix = "arXiv",
	primaryClass = "hep-th",
	doi = "10.21468/SciPostPhys.6.3.039",
	journal = "SciPost Phys.",
	volume = "6",
	number = "3",
	pages = "039",
	year = "2019"
}

@article{Dimofte:2017tpi,
	author = "Dimofte, Tudor and Gaiotto, Davide and Paquette, Natalie M.",
	title = "{Dual boundary conditions in 3d SCFT\textquoteright{}s}",
	eprint = "1712.07654",
	archivePrefix = "arXiv",
	primaryClass = "hep-th",
	doi = "10.1007/JHEP05(2018)060",
	journal = "JHEP",
	volume = "05",
	pages = "060",
	year = "2018"
}

@article{Yoshida:2014ssa,
	author = "Yoshida, Yutaka and Sugiyama, Katsuyuki",
	title = "{Localization of three-dimensional $\mathcal{N}=2$ supersymmetric theories on $S^1 \times D^2$}",
	eprint = "1409.6713",
	archivePrefix = "arXiv",
	primaryClass = "hep-th",
	reportNumber = "KIAS-P14056",
	doi = "10.1093/ptep/ptaa136",
	journal = "PTEP",
	volume = "2020",
	number = "11",
	pages = "113B02",
	year = "2020"
}

@article{Chung:2023qth,
	author = "Chung, Hee-Joong",
	title = "{3d-3d correspondence and 2d $\mathcal{N}$ = (0, 2) boundary conditions}",
	eprint = "2307.10125",
	archivePrefix = "arXiv",
	primaryClass = "hep-th",
	doi = "10.1007/JHEP03(2024)085",
	journal = "JHEP",
	volume = "03",
	pages = "085",
	year = "2024"
}

@article{Chung:2019khu,
	author = "Chung, Hee-Joong",
	title = "{Index for a Model of 3d-3d Correspondence for Plumbed 3-Manifolds}",
	eprint = "1912.13486",
	archivePrefix = "arXiv",
	primaryClass = "hep-th",
	doi = "10.1016/j.nuclphysb.2021.115361",
	journal = "Nucl. Phys. B",
	volume = "965",
	pages = "115361",
	year = "2021"
}

@article{Beem:2023dub,
	author = "Beem, Christopher and Ferrari, Andrea E. V.",
	title = "{Free field realisation of boundary vertex algebras for Abelian gauge theories in three dimensions}",
	eprint = "2304.11055",
	archivePrefix = "arXiv",
	primaryClass = "hep-th",
	month = "4",
	year = "2023"
}

@article{Costello:2018swh,
	author = "Costello, Kevin and Creutzig, Thomas and Gaiotto, Davide",
	title = "{Higgs and Coulomb branches from vertex operator algebras}",
	eprint = "1811.03958",
	archivePrefix = "arXiv",
	primaryClass = "hep-th",
	doi = "10.1007/JHEP03(2019)066",
	journal = "JHEP",
	volume = "03",
	pages = "066",
	year = "2019"
}

@article{Ferrari:2023fez,
	author = "Ferrari, Andrea E. V. and Garner, Niklas and Kim, Heeyeon",
	title = "{Boundary vertex algebras for 3d $\mathcal{N}=4$ rank-0 SCFTs}",
	eprint = "2311.05087",
	archivePrefix = "arXiv",
	primaryClass = "hep-th",
	month = "11",
	year = "2023"
}

@article{Costello:2018fnz,
	author = "Costello, Kevin and Gaiotto, Davide",
	title = "{Vertex Operator Algebras and 3d $ \mathcal{N} $ = 4 gauge theories}",
	eprint = "1804.06460",
	archivePrefix = "arXiv",
	primaryClass = "hep-th",
	doi = "10.1007/JHEP05(2019)018",
	journal = "JHEP",
	volume = "05",
	pages = "018",
	year = "2019"
}

@article{Costello:2020ndc,
	author = "Costello, Kevin and Dimofte, Tudor and Gaiotto, Davide",
	title = "{Boundary Chiral Algebras and Holomorphic Twists}",
	eprint = "2005.00083",
	archivePrefix = "arXiv",
	primaryClass = "hep-th",
	doi = "10.1007/s00220-022-04599-0",
	journal = "Commun. Math. Phys.",
	volume = "399",
	number = "2",
	pages = "1203--1290",
	year = "2023"
}

@article{Cheng:2022rqr,
	author = "Cheng, Miranda C. N. and Chun, Sungbong and Feigin, Boris and Ferrari, Francesca and Gukov, Sergei and Harrison, Sarah M. and Passaro, Davide",
	title = "{3-Manifolds and VOA Characters}",
	eprint = "2201.04640",
	archivePrefix = "arXiv",
	primaryClass = "hep-th",
	doi = "10.1007/s00220-023-04889-1",
	journal = "Commun. Math. Phys.",
	volume = "405",
	number = "2",
	pages = "44",
	year = "2024"
}

@article{Cheng:2018vpl,
	author = "Cheng, Miranda C. N. and Chun, Sungbong and Ferrari, Francesca and Gukov, Sergei and Harrison, Sarah M.",
	title = "{3d Modularity}",
	eprint = "1809.10148",
	archivePrefix = "arXiv",
	primaryClass = "hep-th",
	reportNumber = "CALT-TH-2018-037",
	doi = "10.1007/JHEP10(2019)010",
	journal = "JHEP",
	volume = "10",
	pages = "010",
	year = "2019"
}

@article{Feigin:2018bkf,
	author = "Feigin, Boris and Gukov, Sergei",
	title = "{VOA[$M_4$]}",
	eprint = "1806.02470",
	archivePrefix = "arXiv",
	primaryClass = "hep-th",
	doi = "10.1063/1.5100059",
	journal = "J. Math. Phys.",
	volume = "61",
	number = "1",
	pages = "012302",
	year = "2020"
}

@article{Beem:2014kka,
	author = "Beem, Christopher and Rastelli, Leonardo and van Rees, Balt C.",
	title = "{$ \mathcal{W} $ symmetry in six dimensions}",
	eprint = "1404.1079",
	archivePrefix = "arXiv",
	primaryClass = "hep-th",
	reportNumber = "CERN-PH-TH-2014-056, CERN-PH-TH-2014-56",
	doi = "10.1007/JHEP05(2015)017",
	journal = "JHEP",
	volume = "05",
	pages = "017",
	year = "2015"
}

@article{Beem:2013sza,
	author = "Beem, Christopher and Lemos, Madalena and Liendo, Pedro and Peelaers, Wolfger and Rastelli, Leonardo and van Rees, Balt C.",
	title = "{Infinite Chiral Symmetry in Four Dimensions}",
	eprint = "1312.5344",
	archivePrefix = "arXiv",
	primaryClass = "hep-th",
	reportNumber = "YITP-SB-13-45, CERN-PH-TH-2013-311, HU-EP-13-78",
	doi = "10.1007/s00220-014-2272-x",
	journal = "Commun. Math. Phys.",
	volume = "336",
	number = "3",
	pages = "1359--1433",
	year = "2015"
}

@article{Gang:2019uay,
	author = "Gang, Dongmin and Kim, Nakwoo and Pando Zayas, Leopoldo A.",
	title = "{Precision Microstate Counting for the Entropy of Wrapped M5-branes}",
	eprint = "1905.01559",
	archivePrefix = "arXiv",
	primaryClass = "hep-th",
	reportNumber = "LCTP-19-09",
	doi = "10.1007/JHEP03(2020)164",
	journal = "JHEP",
	volume = "03",
	pages = "164",
	year = "2020"
}

@article{Benini:2019dyp,
	author = "Benini, Francesco and Gang, Dongmin and Pando Zayas, Leopoldo A.",
	title = "{Rotating Black Hole Entropy from M5 Branes}",
	eprint = "1909.11612",
	archivePrefix = "arXiv",
	primaryClass = "hep-th",
	reportNumber = "LCTP-19-24, SISSA 27/2019/FISI",
	doi = "10.1007/JHEP03(2020)057",
	journal = "JHEP",
	volume = "03",
	pages = "057",
	year = "2020"
}

@article{Gang:2018hjd,
	author = "Gang, Dongmin and Kim, Nakwoo",
	title = "{Large $N$ twisted partition functions in 3d-3d correspondence and Holography}",
	eprint = "1808.02797",
	archivePrefix = "arXiv",
	primaryClass = "hep-th",
	doi = "10.1103/PhysRevD.99.021901",
	journal = "Phys. Rev. D",
	volume = "99",
	number = "2",
	pages = "021901",
	year = "2019"
}

@article{Belavin:1984vu,
    author = "Belavin, A. A. and Polyakov, Alexander M. and Zamolodchikov, A. B.",
    editor = "Khalatnikov, I. M. and Mineev, V. P.",
    title = "{Infinite Conformal Symmetry in Two-Dimensional Quantum Field Theory}",
    reportNumber = "CERN-TH-3827",
    doi = "10.1016/0550-3213(84)90052-X",
    journal = "Nucl. Phys. B",
    volume = "241",
    pages = "333--380",
    year = "1984"
}

@article{Alday:2017yxk,
	author = "Alday, Luis Fernando and Benetti Genolini, Pietro and Bullimore, Mathew and van Loon, Mark",
	title = "{Refined 3d-3d Correspondence}",
	eprint = "1702.05045",
	archivePrefix = "arXiv",
	primaryClass = "hep-th",
	doi = "10.1007/JHEP04(2017)170",
	journal = "JHEP",
	volume = "04",
	pages = "170",
	year = "2017"
}

@article{Imamura:2011su,
	author         = "Imamura, Yosuke and Yokoyama, Shuichi",
	title          = "{Index for three dimensional superconformal field
	theories with general R-charge assignments}",
	journal        = "JHEP",
	volume         = "04",
	year           = "2011",
	pages          = "007",
	doi            = "10.1007/JHEP04(2011)007",
	eprint         = "1101.0557",
	archivePrefix  = "arXiv",
	primaryClass   = "hep-th",
	reportNumber   = "UT-11-01, TIT-HEP-607",
	SLACcitation   = "%%CITATION = ARXIV:1101.0557;%%"
}

@article{Kim:2009wb,
	author         = "Kim, Seok",
	title          = "{The Complete superconformal index for N=6 Chern-Simons
	theory}",
	journal        = "Nucl. Phys.",
	volume         = "B821",
	year           = "2009",
	pages          = "241-284",
	doi            = "10.1016/j.nuclphysb.2012.07.015,
	10.1016/j.nuclphysb.2009.06.025",
	note           = "[Erratum: Nucl. Phys.B864,884(2012)]",
	eprint         = "0903.4172",
	archivePrefix  = "arXiv",
	primaryClass   = "hep-th",
	reportNumber   = "IMPERIAL-TP-09-SK-01",
	SLACcitation   = "%%CITATION = ARXIV:0903.4172;%%"
}

@article{Nishioka:2011dq,
	author = "Nishioka, Tatsuma and Tachikawa, Yuji and Yamazaki, Masahito",
	title = "{3d Partition Function as Overlap of Wavefunctions}",
	eprint = "1105.4390",
	archivePrefix = "arXiv",
	primaryClass = "hep-th",
	reportNumber = "PUPT-2376",
	doi = "10.1007/JHEP08(2011)003",
	journal = "JHEP",
	volume = "08",
	pages = "003",
	year = "2011"
}

@article{BGK:2025,
	author = "Baek, Seungjoo and Gang, Dongmin and Kang, Heesu",
	title = {Non-hyperbolic 3-manifolds, 3D rank-0 SCFTs and supersymmetric minimal models},
	journal = "",
	year = "",
	note = "work in progress",
}

@article{turaev:1992modular,
	title={Modular categories and 3-manifold invariants},
	author={Turaev, Vladimir G},
	journal={International Journal of Modern Physics B},
	volume={6},
	number={11n12},
	pages={1807--1824},
	year={1992},
	publisher={World Scientific}
}

@article{Moore:1989vd,
	title={Classical and quantum conformal field theory},
	author={Moore, Gregory and Seiberg, Nathan},
	journal={Communications in Mathematical Physics},
	volume={123},
	pages={177--254},
	year={1989},
	publisher={Springer}
}

@article{Dedushenko:2023cvd,
	author = "Dedushenko, Mykola",
	title = "{On the 4d/3d/2d view of the SCFT/VOA correspondence}",
	eprint = "2312.17747",
	archivePrefix = "arXiv",
	primaryClass = "hep-th",
	month = "12",
	year = "2023"
}

@article{Dedushenko:2018bpp,
	author = "Dedushenko, Mykola and Gukov, Sergei and Nakajima, Hiraku and Pei, Du and Ye, Ke",
	title = "{3d TQFTs from Argyres\textendash{}Douglas theories}",
	eprint = "1809.04638",
	archivePrefix = "arXiv",
	primaryClass = "hep-th",
	reportNumber = "CALT-TH-2018-033",
	doi = "10.1088/1751-8121/abb481",
	journal = "J. Phys. A",
	volume = "53",
	number = "43",
	pages = "43LT01",
	year = "2020"
}

@article{Gang:2022kpe,
	author = "Gang, Dongmin and Kim, Dongyeob",
	title = "{Generalized non-unitary Haagerup-Izumi modular data from 3D S-fold SCFTs}",
	eprint = "2211.13561",
	archivePrefix = "arXiv",
	primaryClass = "hep-th",
	doi = "10.1007/JHEP03(2023)185",
	journal = "JHEP",
	volume = "03",
	pages = "185",
	year = "2023"
}

@article{Gang:2023ggt,
	author = "Gang, Dongmin and Kim, Dongyeob and Lee, Sungjay",
	title = "{A non-unitary bulk-boundary correspondence: Non-unitary Haagerup RCFTs from S-fold SCFTs}",
	eprint = "2310.14877",
	archivePrefix = "arXiv",
	primaryClass = "hep-th",
	reportNumber = "KIAS-P23052",
	month = "10",
	year = "2023"
}

@article{Baek:2024tuo,
	author = "Baek, Seungjoo and Gang, Dongmin",
	title = "{3D bulk field theories for 2D non-unitary N=1 supersymmetric minimal models}",
	eprint = "2405.05746",
	archivePrefix = "arXiv",
	primaryClass = "hep-th",
	month = "5",
	year = "2024"
}

@article{Benini:2010uu,
	author = "Benini, Francesco and Tachikawa, Yuji and Xie, Dan",
	title = "{Mirrors of 3d Sicilian theories}",
	eprint = "1007.0992",
	archivePrefix = "arXiv",
	primaryClass = "hep-th",
	reportNumber = "MIFPA-10-27, PUTP-2344",
	doi = "10.1007/JHEP09(2010)063",
	journal = "JHEP",
	volume = "09",
	pages = "063",
	year = "2010"
}

@article{Gukov:2016gkn,
	author = "Gukov, Sergei and Putrov, Pavel and Vafa, Cumrun",
	title = "{Fivebranes and 3-manifold homology}",
	eprint = "1602.05302",
	archivePrefix = "arXiv",
	primaryClass = "hep-th",
	reportNumber = "CALT-2016-004",
	doi = "10.1007/JHEP07(2017)071",
	journal = "JHEP",
	volume = "07",
	pages = "071",
	year = "2017"
}

@article{Gadde:2013sca,
	author = "Gadde, Abhijit and Gukov, Sergei and Putrov, Pavel",
	editor = "Ballmann, Werner and Blohmann, Christian and Faltings, Gerd and Teichner, Peter and Zagier, Don",
	title = "{Fivebranes and 4-manifolds}",
	eprint = "1306.4320",
	archivePrefix = "arXiv",
	primaryClass = "hep-th",
	reportNumber = "CALT-68-2904",
	doi = "10.1007/978-3-319-43648-7_7",
	journal = "Prog. Math.",
	volume = "319",
	pages = "155--245",
	year = "2016"
}

@article{Eckhard:2019jgg,
	author = "Eckhard, Julius and Kim, Heeyeon and Schafer-Nameki, Sakura and Willett, Brian",
	title = "{Higher-Form Symmetries, Bethe Vacua, and the 3d-3d Correspondence}",
	eprint = "1910.14086",
	archivePrefix = "arXiv",
	primaryClass = "hep-th",
	doi = "10.1007/JHEP01(2020)101",
	journal = "JHEP",
	volume = "01",
	pages = "101",
	year = "2020"
}

@article{Chung:2014qpa,
	author = "Chung, Hee-Joong and Dimofte, Tudor and Gukov, Sergei and Sulkowski, Piotr",
	title = "{3d-3d Correspondence Revisited}",
	eprint = "1405.3663",
	archivePrefix = "arXiv",
	primaryClass = "hep-th",
	reportNumber = "CALT-68-2887",
	doi = "10.1007/JHEP04(2016)140",
	journal = "JHEP",
	volume = "04",
	pages = "140",
	year = "2016"
}

@article{Bonetti:2024cvq,
	author = "Bonetti, Federico and Schafer-Nameki, Sakura and Wu, Jingxiang",
	title = "{MTC$[M_3, G]$: 3d Topological Order Labeled by Seifert Manifolds}",
	eprint = "2403.03973",
	archivePrefix = "arXiv",
	primaryClass = "hep-th",
	month = "3",
	year = "2024"
}

@article{Comi:2023lfm,
	author = "Comi, Riccardo and Harding, William and Mekareeya, Noppadol",
	title = "{Chern-Simons-Trinion theories: One-form symmetries and superconformal indices}",
	eprint = "2305.07055",
	archivePrefix = "arXiv",
	primaryClass = "hep-th",
	doi = "10.1007/JHEP09(2023)060",
	journal = "JHEP",
	volume = "09",
	pages = "060",
	year = "2023"
}

@article{Gukov:2017kmk,
	author = "Gukov, Sergei and Pei, Du and Putrov, Pavel and Vafa, Cumrun",
	title = "{BPS spectra and 3-manifold invariants}",
	eprint = "1701.06567",
	archivePrefix = "arXiv",
	primaryClass = "hep-th",
	reportNumber = "CALT-TH-2016-039",
	doi = "10.1142/S0218216520400039",
	journal = "J. Knot Theor. Ramifications",
	volume = "29",
	number = "02",
	pages = "2040003",
	year = "2020"
}

@article{Gang:2023rei,
	author = "Gang, Dongmin and Kim, Heeyeon and Stubbs, Spencer",
	title = "{Three-Dimensional Topological Field Theories and Nonunitary Minimal Models}",
	eprint = "2310.09080",
	archivePrefix = "arXiv",
	primaryClass = "hep-th",
	doi = "10.1103/PhysRevLett.132.131601",
	journal = "Phys. Rev. Lett.",
	volume = "132",
	number = "13",
	pages = "131601",
	year = "2024"
}

@article{dunfield:2018census,
	title={A census of exceptional Dehn fillings},
	author={Dunfield, Nathan M},
	journal={Characters in low-dimensional topology},
	volume={760},
	pages={143--155},
	year={2018}
}

@article{Cui:2021lyi,
	author = "Cui, Shawn X. and Qiu, Yang and Wang, Zhenghan",
	title = "{From Three Dimensional Manifolds to Modular Tensor Categories}",
	eprint = "2101.01674",
	archivePrefix = "arXiv",
	primaryClass = "math.QA",
	doi = "10.1007/s00220-022-04517-4",
	journal = "Commun. Math. Phys.",
	volume = "397",
	number = "3",
	pages = "1191--1235",
	year = "2023"
}

@article{Choi:2022dju,
	author = "Choi, Sunjin and Gang, Dongmin and Kim, Hee-Cheol",
	title = "{Infrared phases of 3D class R theories}",
	eprint = "2206.11982",
	archivePrefix = "arXiv",
	primaryClass = "hep-th",
	reportNumber = "KIAS-P22046",
	doi = "10.1007/JHEP11(2022)151",
	journal = "JHEP",
	volume = "11",
	pages = "151",
	year = "2022"
}

@article{Assel:2022row,
	author = "Assel, Benjamin and Tachikawa, Yuji and Tomasiello, Alessandro",
	title = "{On $ \mathcal{N} $ = 4 supersymmetry enhancements in three dimensions}",
	eprint = "2209.13984",
	archivePrefix = "arXiv",
	primaryClass = "hep-th",
	doi = "10.1007/JHEP03(2023)170",
	journal = "JHEP",
	volume = "03",
	pages = "170",
	year = "2023"
}

@article{Closset:2016arn,
	author = "Closset, Cyril and Kim, Heeyeon",
	title = "{Comments on twisted indices in 3d supersymmetric gauge theories}",
	eprint = "1605.06531",
	archivePrefix = "arXiv",
	primaryClass = "hep-th",
	doi = "10.1007/JHEP08(2016)059",
	journal = "JHEP",
	volume = "08",
	pages = "059",
	year = "2016"
}

@article{Benini:2016hjo,
	author = "Benini, Francesco and Zaffaroni, Alberto",
	editor = "Li, Si and Lian, Bong H. and Song, Wei and Yau, Shing-Tung",
	title = "{Supersymmetric partition functions on Riemann surfaces}",
	eprint = "1605.06120",
	archivePrefix = "arXiv",
	primaryClass = "hep-th",
	reportNumber = "SISSA-28-2016-FISI",
	journal = "Proc. Symp. Pure Math.",
	volume = "96",
	pages = "13--46",
	year = "2017"
}

@article{Benini:2015noa,
	author = "Benini, Francesco and Zaffaroni, Alberto",
	title = "{A topologically twisted index for three-dimensional supersymmetric theories}",
	eprint = "1504.03698",
	archivePrefix = "arXiv",
	primaryClass = "hep-th",
	reportNumber = "IMPERIAL-TP-2015-FB-01",
	doi = "10.1007/JHEP07(2015)127",
	journal = "JHEP",
	volume = "07",
	pages = "127",
	year = "2015"
}

@article{Closset:2018ghr,
	author = "Closset, Cyril and Kim, Heeyeon and Willett, Brian",
	title = "{Seifert fibering operators in 3d $\mathcal{N}=2$ theories}",
	eprint = "1807.02328",
	archivePrefix = "arXiv",
	primaryClass = "hep-th",
	reportNumber = "CERN-TH-2018-156",
	doi = "10.1007/JHEP11(2018)004",
	journal = "JHEP",
	volume = "11",
	pages = "004",
	year = "2018"
}

@article{Witten:1988hf,
    author = "Witten, Edward",
    editor = "Mitra, Asoke N.",
    title = "{Quantum Field Theory and the Jones Polynomial}",
    reportNumber = "IASSNS-HEP-88-33",
    doi = "10.1007/BF01217730",
    journal = "Commun. Math. Phys.",
    volume = "121",
    pages = "351--399",
    year = "1989"
}

@article{Pei:2015jsa,
	author = "Pei, Du and Ye, Ke",
	title = "{A 3d-3d appetizer}",
	eprint = "1503.04809",
	archivePrefix = "arXiv",
	primaryClass = "hep-th",
	reportNumber = "CALT-TH-2015-013",
	doi = "10.1007/JHEP11(2016)008",
	journal = "JHEP",
	volume = "11",
	pages = "008",
	year = "2016"
}

@article{Gadde:2013wq,
    author = "Gadde, Abhijit and Gukov, Sergei and Putrov, Pavel",
    title = "{Walls, Lines, and Spectral Dualities in 3d Gauge Theories}",
    eprint = "1302.0015",
    archivePrefix = "arXiv",
    primaryClass = "hep-th",
    reportNumber = "CALT-68-2905",
    doi = "10.1007/JHEP05(2014)047",
    journal = "JHEP",
    volume = "05",
    pages = "047",
    year = "2014"
}

@article{Gang:2021hrd,
	author = "Gang, Dongmin and Kim, Sungjoon and Lee, Kimyeong and Shim, Myungbo and Yamazaki, Masahito",
	title = "{Non-unitary TQFTs from 3D $ \mathcal{N} $ = 4 rank 0 SCFTs}",
	eprint = "2103.09283",
	archivePrefix = "arXiv",
	primaryClass = "hep-th",
	doi = "10.1007/JHEP08(2021)158",
	journal = "JHEP",
	volume = "08",
	pages = "158",
	year = "2021"
}

@article{Cho:2020ljj,
	author = "Cho, Gil Young and Gang, Dongmin and Kim, Hee-Cheol",
	title = "{M-theoretic Genesis of Topological Phases}",
	eprint = "2007.01532",
	archivePrefix = "arXiv",
	primaryClass = "hep-th",
	doi = "10.1007/JHEP11(2020)115",
	journal = "JHEP",
	volume = "11",
	pages = "115",
	year = "2020"
}

@article{Gang:2019jut,
	author = "Gang, Dongmin and Yamazaki, Masahito",
	title = "{Expanding 3d $ \mathcal{N} $ = 2 theories around the round sphere}",
	eprint = "1912.09617",
	archivePrefix = "arXiv",
	primaryClass = "hep-th",
	reportNumber = "IPMU19-018",
	doi = "10.1007/JHEP02(2020)102",
	journal = "JHEP",
	volume = "02",
	pages = "102",
	year = "2020"
}

@article{Gang:2018huc,
	author = "Gang, Dongmin and Yamazaki, Masahito",
	title = "{Three-dimensional gauge theories with supersymmetry enhancement}",
	eprint = "1806.07714",
	archivePrefix = "arXiv",
	primaryClass = "hep-th",
	reportNumber = "IPMU18-0081",
	doi = "10.1103/PhysRevD.98.121701",
	journal = "Phys. Rev. D",
	volume = "98",
	number = "12",
	pages = "121701",
	year = "2018"
}

@article{Faddeev:1993rs,
    author = "Faddeev, L. D. and Kashaev, R. M.",
    title = "{Quantum Dilogarithm}",
    eprint = "hep-th/9310070",
    archivePrefix = "arXiv",
    reportNumber = "HU-TFT-93-56",
    doi = "10.1142/S0217732394000447",
    journal = "Mod. Phys. Lett. A",
    volume = "9",
    pages = "427--434",
    year = "1994"
}

@article{Garozzo:2019ejm,
	author         = "Garozzo, Ivan and Lo Monaco, Gabriele and Mekareeya,
	Noppadol and Sacchi, Matteo",
	title          = "{Supersymmetric Indices of 3d S-fold SCFTs}",
	year           = "2019",
	eprint         = "1905.07183",
	archivePrefix  = "arXiv",
	primaryClass   = "hep-th",
	SLACcitation   = "%%CITATION = ARXIV:1905.07183;%%"
}

@article{Dimofte:2011py,
      author         = "Dimofte, Tudor and Gaiotto, Davide and Gukov, Sergei",
      title          = "{3-Manifolds and 3d Indices}",
      journal        = "Adv. Theor. Math. Phys.",
      volume         = "17",
      year           = "2013",
      number         = "5",
      pages          = "975-1076",
      doi            = "10.4310/ATMP.2013.v17.n5.a3",
      eprint         = "1112.5179",
      archivePrefix  = "arXiv",
      primaryClass   = "hep-th",
      SLACcitation   = "%%CITATION = ARXIV:1112.5179;%%"
}

@article{Gaiotto:2008ak,
      author         = "Gaiotto, Davide and Witten, Edward",
      title          = "{S-Duality of Boundary Conditions In N=4 Super Yang-Mills
                        Theory}",
      journal        = "Adv. Theor. Math. Phys.",
      volume         = "13",
      year           = "2009",
      number         = "3",
      pages          = "721-896",
      doi            = "10.4310/ATMP.2009.v13.n3.a5",
      eprint         = "0807.3720",
      archivePrefix  = "arXiv",
      primaryClass   = "hep-th",
      SLACcitation   = "%%CITATION = ARXIV:0807.3720;%%"
}

@article{Terashima:2011qi,
      author         = "Terashima, Yuji and Yamazaki, Masahito",
      title          = "{SL(2,R) Chern-Simons, Liouville, and Gauge Theory on
                        Duality Walls}",
      journal        = "JHEP",
      volume         = "08",
      year           = "2011",
      pages          = "135",
      doi            = "10.1007/JHEP08(2011)135",
      eprint         = "1103.5748",
      archivePrefix  = "arXiv",
      primaryClass   = "hep-th",
      reportNumber   = "PUPT-2368",
      SLACcitation   = "%%CITATION = ARXIV:1103.5748;%%"
}

@article{Gang:2018wek,
      author         = "Gang, Dongmin and Yonekura, Kazuya",
      title          = "{Symmetry enhancement and closing of knots in 3d/3d
                        correspondence}",
      year           = "2018",
      eprint         = "1803.04009",
      archivePrefix  = "arXiv",
      primaryClass   = "hep-th",
      reportNumber   = "IPMU-18-0045",
      SLACcitation   = "%%CITATION = ARXIV:1803.04009;%%"
}

@article{Hama:2010av,
      author         = "Hama, Naofumi and Hosomichi, Kazuo and Lee, Sungjay",
      title          = "{Notes on SUSY Gauge Theories on Three-Sphere}",
      journal        = "JHEP",
      volume         = "03",
      year           = "2011",
      pages          = "127",
      doi            = "10.1007/JHEP03(2011)127",
      eprint         = "1012.3512",
      archivePrefix  = "arXiv",
      primaryClass   = "hep-th",
      reportNumber   = "DAMTP-2010-129, YITP-10-100",
      SLACcitation   = "%%CITATION = ARXIV:1012.3512;%%"
}

@article{Jafferis:2010un,
      author         = "Jafferis, Daniel L.",
      title          = "{The Exact Superconformal R-Symmetry Extremizes Z}",
      journal        = "JHEP",
      volume         = "05",
      year           = "2012",
      pages          = "159",
      doi            = "10.1007/JHEP05(2012)159",
      eprint         = "1012.3210",
      archivePrefix  = "arXiv",
      primaryClass   = "hep-th",
      SLACcitation   = "%%CITATION = ARXIV:1012.3210;%%"
}

@article{Kapustin:2009kz,
      author         = "Kapustin, Anton and Willett, Brian and Yaakov, Itamar",
      title          = "{Exact Results for Wilson Loops in Superconformal
                        Chern-Simons Theories with Matter}",
      journal        = "JHEP",
      volume         = "03",
      year           = "2010",
      pages          = "089",
      doi            = "10.1007/JHEP03(2010)089",
      eprint         = "0909.4559",
      archivePrefix  = "arXiv",
      primaryClass   = "hep-th",
      reportNumber   = "CALT-68-2750",
      SLACcitation   = "%%CITATION = ARXIV:0909.4559;%%"
}

@article{Dimofte:2011ju,
      author         = "Dimofte, Tudor and Gaiotto, Davide and Gukov, Sergei",
      title          = "{Gauge Theories Labelled by Three-Manifolds}",
      journal        = "Commun. Math. Phys.",
      volume         = "325",
      year           = "2014",
      pages          = "367-419",
      doi            = "10.1007/s00220-013-1863-2",
      eprint         = "1108.4389",
      archivePrefix  = "arXiv",
      primaryClass   = "hep-th",
      reportNumber   = "CALT-68-2847",
      SLACcitation   = "%%CITATION = ARXIV:1108.4389;%%"
}
	
\end{document}